\documentclass[pre,twocolumn,showpacs]{revtex4}

\usepackage{epsfig}

\begin{document}

\title{Diffusion of Tagged Particle in an Exclusion Process}
\author{E. Barkai }
\affiliation{ Department of Physics, Bar Ilan University, Ramat-Gan
52900, Israel}
\author{R. Silbey }

\affiliation{Department of Chemistry, Massachusetts Institute of Technology, Cambridge, Massachusetts 02139, USA} 

%%%%%%%%%%%%%%%%%%%%%%%%%%%%%%%%%%%%%%%%%%%%%%%%%%%%%%%%%%%%%%%%%%%%%%%%%%%%%%%
%
% A B S T R A C T 
%
%%%%%%%%%%%%%%%%%%%%%%%%%%%%%%%%%%%%%%%%%%%%%%%%%%%%%%%%%%%%%%%%%%%%%%%%%%%%%%%
\begin{abstract}

 We study the diffusion of tagged hard core interacting particles
under the influence of an external force field. 
Using the Jepsen line we map this many particle problem onto a single
particle one. 
We obtain general
equations for the distribution and the
 mean square displacement $\langle (x_T)^2 \rangle$
of the tagged center particle
valid for rather general external force  fields and initial conditions. 
A wide range of physical behaviors emerge which are 
very different
than the classical single file sub-diffusion 
$\langle (x_T)^2 \rangle \sim t^{1/2}$ found for uniformly 
distributed particles in an infinite space and
in the absence of force fields. 
For symmetric initial conditions and potential fields we find
$\langle (x_T)^2 \rangle = { {\cal R} \left( 1 - {\cal R} \right)\over 2 N {\it r} ^2 } $ where $2 N$ is the (large) number of particles in the
system, ${\cal R}$ is a single particle reflection coefficient obtained
from the single particle Green function and initial conditions, and
$r$ its derivative. 
We show that this equation is related to the mathematical theory of 
order statistics and it can be used to find $\langle (x_T)^2 \rangle$ 
even when the motion between collision events is not Brownian (e.g. it might
be ballistic, or anomalous diffusion). As an example we derive
the Percus relation for non Gaussian diffusion. 

\end{abstract}

% these pacs are taken from our prl

\maketitle

%	05.40.Fb; 02.50.Ey; 05.10.Gg

\section{Introduction}

Systems of particles governed by stochastic dynamics and exclusion interactions
have been studied for a long time \cite{Ligget,Mukamel,Derrida}.
One aspect of this problem is the motion of a tagged particle sometimes called
the tracer particle \cite{Ha,Jepsen,Le,alexander}. 
The diffusion of a tagged particle,
in a  one dimensional system of Brownian particles, interacting
via hard core interaction is a model for the motion of a single
molecule in a crowded one dimensional environment such as  a biological
pore or channel \cite{karger,Burada,Kim,Golan}, and 
in experimentally studied physical systems
such as zeolites \cite{Hahn1}, confined colloid particles 
\cite{Wei,Lutz,LMCRD} and charged spheres in circular channels \cite{Saint}.
Since particles do not pass each other such diffusion processes are
called single file diffusion.

 Confinement of a tagged Brownian particle, due to its interaction with other
Brownian particles, leads to a slow-down of the diffusion of the
tagged particle $\langle (x_T)^2 \rangle \propto t^{1/2} $ instead of normal
diffusion $\langle (x_T)^2 \rangle \propto t$ when  the particles
in the system are uniformly distributed \cite{Ha,Le}. 
Such many body problems
can be treated using methods  
which exploit the relation between the dynamics of the interacting tagged
particle and the motion of a particle  free of interactions
 \cite{Jepsen,Le,Percus,Percus1,Leb,Kalinay}. In recent years
at-least six new directions  of research emerged. (i) The effect
of an external field acting on all the particles 
\cite{Silbey}
or on the tagged particle only
\cite{Oshanin,Lizana} 
has attracted attention since pores induce entropic barriers \cite{Burada}
and are generally
inhomogeneous. In this category the examples
of  single file motion in a periodic potential \cite{Mar}
or in a box \cite{LIAM}
 were considered in detail. (ii) Initial conditions may have
a profound impact on diffusion of the tagged particle. For example
if particles start as a narrow Gaussian packet
the diffusion of the tagged center particle  increases linearly in time
$\langle (x_T)^2 \rangle\propto
t$ \cite{Aslangul} (see details below). 
Power law initial conditions induce $\langle (x_T)^2 \rangle \propto t^\xi$
and $\xi$ is  neither $1$ nor $1/2$ \cite{Ophir} 
(see further details in text). 
When an external
field is acting on the system, it is important to consider  
 non uniform initial conditions where 
the density of particles
is determined naturally from Boltzmann's distribution. 
(iii) In some investigations
the underlying motion is not normal diffusion, instead the particles  may be non-Brownian,
and following anomalous kinetics \cite{Ophir,Band}. (iv) Usually hard core interactions are considered,
although the hard problem  of more general interactions has been recently 
treated in \cite{Lizana,kollmann,peeters}.
Screened hydrodynamic interactions which seem important at short times
were investigated in \cite{Cui} both theoretically and experimentally.
Granular single file diffusion with inelastic collisions 
shows the typical $t^{1/2}$ sub-diffusion \cite{Cecconi,Vulp}.
(v)  Interacting particles in systems with quenched disorder is yet another 
challenge.
The motion of a tagged particle was recently treated in the context of single file diffusion in the Sinai model
\cite{Ben}. (vi) Finally, if the motion of the tagged particle is not
normal Brownian motion, what stochastic theory replaces the usual 
Brownian-Langevin framework? In this direction interesting connections
to fractional calculus emerged \cite{Lizana,taloni,Lim}. 
Roughly speaking and under certain
conditions half order time derivatives $(d^{1/2} / d t^{1/2})$
enter in the Langevin equation \cite{Lizana}
and fractional Brownian noise replaces white noise. 

 Here we provide a very general theory of single file diffusion of the center tagged particle,
valid in the presence or the absence of an external potential field $V(x)$, as
well as for thermal and non-thermal initial conditions. Our main results
reproduce previously obtained formulas and many new ones, by mapping the
many particle problem onto a single particle model. Our  method,
explained in Sec. \ref{SecMethod}, exploits the 
theoretical concept of the Jepsen line \cite{Jepsen}, 
is limited to hard core point particles,
but, as we show in Sec. \ref{SecBeyond}, is not limited to Brownian particles.
After providing general results in Sec. \ref{SecRes} 
we limit our  attention to symmetric
initial conditions and potentials where the tagged particle has no
average drift. A general
relation between the mean square displacement
of the tagged particle and reflection probability of the non interacting
particle is given in Eq.  
(\ref{eq21}).
Detailed calculations of the mean square displacement 
of the tagged particle then
follow, for special choices of
force fields and initial conditions, 
in Sec. \ref{secPI}.
  As we discuss  in sub-section 
\ref{SecGumbel},
in certain limits our problem is related
to order statistics \cite{Gumbel}, a fact worth mentioning since it allows
us to solve our problem and related ones using known methods.
In Sec. 
\ref{SecBeyond}, we discuss non-Brownian kinds of motion and the Percus
relation. 
A brief report of part of our results was
recently published \cite{Silbey}.

\section{Model and Methods}
\label{SecMethod} 

 In our model,
 $2 N + 1$ identical point particles with hard core particle-particle
interactions are undergoing Brownian motion in one dimension,
so particles cannot
pass each other. The diffusion constant of particles free of interaction
is $D$. As mentioned, 
 towards the end of the paper we discuss the more general case
where the dynamics between collision events is not necessarily Brownian. 
 An external potential
$V(x)$ acts on the particles.  The system stretches from 
$-\overline{L}$ to $\overline{L}$; however, unless stated otherwise 
we will let $\overline{L} \to \infty$ and obtain a thermodynamic 
limit where $N /\overline{L}$ is fixed.
We tag the center particle, which clearly has $N$
particles to its left and $N$ to it right. Initially the tagged particle is
at the origin $x=0$. The motion of a single particle in the absence
of interactions with other particles is described by a single
particle Green function $g(x,x_0,t)$, with the initial conditions
$g(x,x_0,0)= \delta(x - x_0)$. In the case of over damped Brownian
motion the Green function is the solution of the Fokker-Planck equation
\cite{Risken} 
\begin{equation}
{ \partial g(x,x_0,t) \over \partial t} = D \left[ {\partial^2 \over \partial x^2} - {1 \over k_b T} {\partial \over \partial x} F(x) \right] g(x,x_0, t),
\label{eqzzz}
\end{equation}
where $F(x) = - V'(x)$ is the force field, $T$ is the temperature and
$k_b$ is Boltzmann's constant.  
The initial conditions of $N$ particles residing initially to the
right (left)  of the test  particle are drawn from
the probability density function (PDF)
$f_R(x_0)$ $(f_L(x_0))$ respectively. We consider
an ensemble of trajectories and average over trajectories
and initial
conditions (see details below). 
Our goal is to obtain the PDF of the position $x_T$ of the
tagged center particle $P(x_T)$  in the large $N$
limit.  

\subsection{The Jepsen Line} 

\begin{figure}
\begin{center}
\epsfig{figure=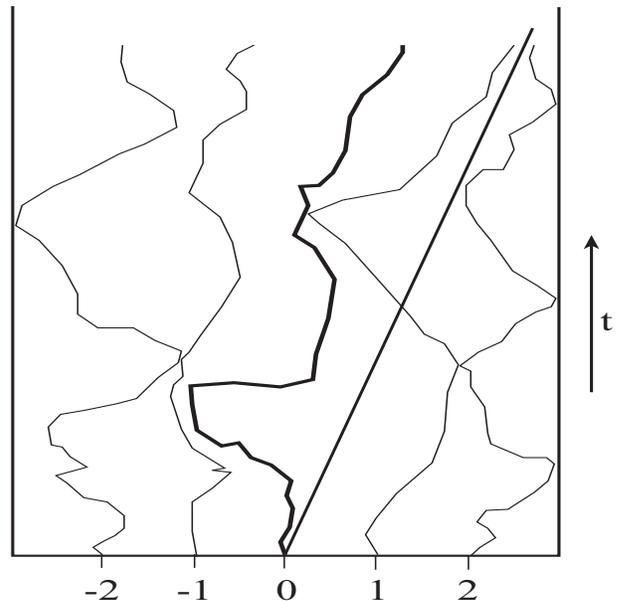,totalheight=0.34\textheight, width=0.45\textwidth}
\end{center}
\caption{ 
Schematic motion of Brownian particles in a box, where particles
cannot penetrate through each other.
 The center tagged particle label is $0$, its
trajectory is restricted by collisions with neighboring particles. The straight
line is the Jepsen line, it follows $v t$ as explained in the  text.
As explained in the text,  in
an equivalent non-interacting picture, we allow particles to pass through
each other, and at time $t$ we search for the position of the particle
which has $N$ particles to its right and $N$ to its left (i.e. the
center particle).  
}
\label{fig1}
\end{figure}

 A schematic diagram of the problem is presented
 in Fig. \ref{fig1} for particles in a box. 
 Initial positions of particles
are given by $x_{0} ^{j}$ where $j=-N, \cdots, 0, \cdots N$ where $j$
 is the label
of the interacting particles (see Fig. \ref{fig1}). The tagged
particle whose coordinate is denoted with $x_T (t)$, is the
center particle $j=0$ (bold line in Fig. \ref{fig1}).
Initially the tagged particle is at the origin $x_T (0) =0$. 
  Since particles do not pass each 
other, their order is clearly maintained, and the number of particles
to the left and right of the tagged particle  $N$ is  fixed. 

In Fig. \ref{fig1} the straight line which starts at $x=0$ is called
the Jepsen line and follows $x(t) = v t$,
where $v$ is a test velocity \cite{Jepsen,Le}. 
We label the interacting particles according to their initial position,
increasing to the right (see Fig. \ref{fig1}). 
The tagged particle starts just to the right
of the Jepsen line so at $t=0$ the label of the particle
to the right of the Jepsen line is zero. 
In this system we
have $2 N + 1$ particles. 
Hence initially we have
$N$ particles to the left of the Jepsen line and including the tagged particle  $N+1$
particles to the right. 

 Let $\tilde{\alpha}(t)$ be the label number of the first particle
situated to the
right of the Jepsen line. According to our rules, at 
$t=0$ we have $\tilde{\alpha} =0$, and then the random variable 
$\tilde{\alpha}$ will increase or
decrease in steps of $+1$ or $-1$ according to:\\
i) if a particle crosses the Jepsen line from left to right $\tilde{\alpha} \to \tilde{\alpha} - 1$ \\
ii) if a particle crosses the Jepsen line from right to left 
$\tilde{\alpha} \to \tilde{\alpha}+  1$.
Thus the counter $\tilde{\alpha}$ is performing a random walk
decreasing or increasing its value $+1$ or $-1$ at random times. 

\begin{figure}
\begin{center}
\epsfig{figure=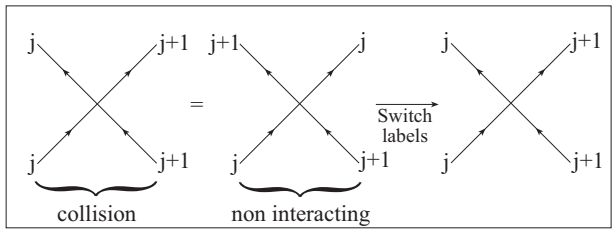, width=0.45\textwidth}
\end{center}
\caption{ 
In one dimension, the paths of a pair of particles in
a hard core collision event can be represented by   
two non interacting particles which pass through each other,
and then their label numbers are switched.  
}
\label{fig2}
\end{figure}

 A collision between two hard core particles is represented schematically in
Fig. \ref{fig2}. In one dimension a hard core 
collision event is equivalent to two  particles
that pass through each other, i.e. non interacting particles,
 and then after the particles cross each other, the labels of
the pair of  particles are switched (see Fig. \ref{fig2}). 
 Instead of relabeling
particles after each collision, we let particles pass through each 
other, and then at time $t$ we label our particles (or if we are
interested only in tagged particle, locate the central particle). 
Operationally this means that in the time interval
$(0,t)$ we view the particles
as non interacting, and then find the particle with $N$ particles
to its right and $N$ to its left, which is equivalent
to  the tagged particle
in the interacting system. 
Hence  the problem is related to the mathematical topic of order statistics \cite{Gumbel}
as we will discuss briefly later. 

  Following Jepsen \cite{Jepsen} and Levitt \cite{Le} we introduce the
stochastic process for the non interacting particles  $\alpha(t)$ where:\\
i) if a particle crosses the Jepsen line from left to right $\alpha \to \alpha - 1$ \\
ii) if a particle crosses the Jepsen line from right to left 
$\alpha \to \alpha+  1$.
The process $\alpha(t)$ 
is the same as the process $\tilde{\alpha}(t)$,
in statistical sense. 

The event where $\alpha(t)=0$ implies that the number of particles
crossing the Jepsen line from right to left is the same
as those crossing from left to right. 
Switching back to the interacting system, the event $\alpha(t)=0$
means  that the particle
to the right of the Jepsen line at time $t=0$ is also
there at time $t$. Hence the probability that $\alpha(t)=0$ is
the same as the probability of finding the tagged interacting
 particle
to be the first particle to  the right of the Jepsen line. 
Therefore we will calculate the probability for $\alpha(t)=0$ and
with it obtain in the large $N$ limit,  the probability $P(x_T) {\rm d} x_T$
of finding
the tagged particle in a small interval $x_T,x_T+{\rm d} x_T$.

The random variable $\alpha$ is a sum of many random variables
\begin{equation}
\alpha = {\sum_{j=-N}^{N}}' \delta \alpha_j 
\label{eq01}
\end{equation}
where $\delta \alpha_j$ is the number of times particle $j$
crossed the Jepsen line from right to left minus the number of
times it crossed the line from left to right. Clearly $\delta \alpha_j$ 
may attain the values $-1$ or $0$ if the particle started on the left,
 or $1$ or $0$ if it started on the right. Since we are interested in the
large $N$ limit we may neglect the contribution of $\delta \alpha_0$,
which is indicated by the prime in the sum \cite{remark0}.

 We calculate the probability
$P_N (\alpha)$ 
 of the random variable $\alpha$. For that we designate $P_{LL}(x_0 ^{-j})$,
the probability that particle $-j$, starting to the left of the 
Jepsen line at $t=0$ at  $x_0 ^{-j}<0$, is found at time
$t$ to the left of the Jepsen line. Clearly for the corresponding trajectory
$\delta \alpha_{-j}=0$, since the particle crossed
the Jepsen line an even number of times, or did not cross it at all. 
Similarly, $P_{LR} (x_0 ^{-j})$ is the probability to start to the
left of the line and end to its right, and $P_{RR}(x_0 ^j)$, $P_{RL}(x_0 ^j)$
are defined similarly for particles starting on $x_0 ^j>0$
to the right of the line.

In our non interacting picture, the motion of particles is
independent, hence we can use random walk theory \cite{Montroll} and 
 Fourier series 
to find $P_N (\alpha)$. 
It is convenient to rewrite Eq. (\ref{eq01}) 
\begin{equation}
\alpha = \sum_{j=1}^N  \Delta \alpha_j 
\label{eq01a}
\end{equation}
where $\Delta \alpha_j = \delta \alpha_j + \delta \alpha_{-j}$ may attain the values $1 , 0, -1$.
Each summand $\Delta \alpha_j$ takes into account 
 one  particle starting to the
left of the Jepsen line and one to the right. 

\begin{figure}
\begin{center}
\epsfig{figure=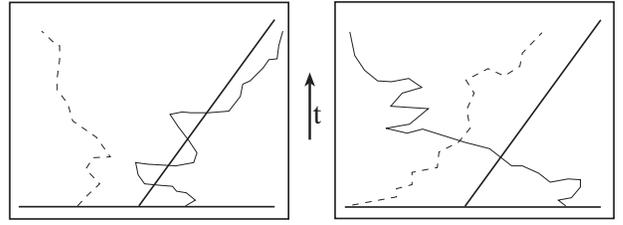, width=0.45\textwidth}
\end{center}
\caption{ 
Trajectories crossing the Jepsen line. In the left panel one particle
started on $L$ and ended on $L$ the other on $R$ and ended on $R$.
These two paths are assigned the probability $P_{LL}(x_0 ^{-1}) P_{RR}
(x_0 ^1)$ and yield in Eq. 
(\ref{eq02})
 $\alpha=0$. Similarly the trajectories on the right panel 
correspond to $P_{LL}(x_0 ^{-1} ) P_{RL} (x_0 ^{1} ) $ and they
contribute $\alpha= 1$.  
}
\label{fig3}
\end{figure}

 First consider $N=1$, that is one particle that starts on the
left of the Jepsen line and one to its right. 
Then $\alpha= \Delta \alpha_1= \delta \alpha_{1} + \delta \alpha_{-1}$ and
as mentioned $\alpha=1$ or $0$ or $-1$.
Examples for possible trajectories are shown in Fig. \ref{fig3}.
It is easy to see that
\begin{equation}
P_{N=1} \left( \alpha \right) = \left\{
\begin{array}{l l}
\alpha = 1 & P_{RL}(x_0 ^1 ) P_{LL} (x_0 ^{-1} ) \\
\alpha = 0 & P_{LL}(x_0 ^{-1} ) P_{RR} (x_0 ^{1} ) + P_{LR}(x_0 ^{-1} ) P_{RL} (x_0 ^{1})  \\
\alpha = -1 & P_{RR}(x_0 ^1 ) P_{LR} (x_0 ^{-1} ) .
\end{array}
\right.
\label{eq02}
\end{equation}
We define the structure function
\begin{widetext}
\begin{equation}
 \lambda(\phi, x_0 ^{-j},x_0 ^j) = 
e^{ i \phi}
P_{LL}(x_0 ^{-j} )
 P_{RL} (x_0 ^j) 
+ \left[ P_{LL} (x_0 ^{-j}) P_{RR} (x_0 ^j ) + P_{LR} (x_0 ^{-j}) P_{RL} (x_0 ^j ) \right] +
 e^{- i \phi} 
P_{LR} (x_0 ^{-j} ) 
P_{RR}( x_0 ^j). 
\label{eq03}
\end{equation}
\end{widetext}
The structure function describes a single step in the random walk
Eq. (\ref{eq01a})
in the following usual way: the coefficient of $\exp( i \phi)$ 
is the probability that $\Delta \alpha_j$ is equal one (i.e. $P_{LL} P_{RL})$,
the coefficient of $\exp( i \phi) =1$ (i.e. $\phi=0$)
yields
the probability $\Delta \alpha_j =0$, and similarly for $\exp( - i \phi)$.
Since the summands in Eq. (\ref{eq01a}) are independent,
Fourier analysis gives:
\begin{equation} 
P_N (\alpha) = {1 \over 2 \pi} \int_{-\pi} ^\pi {\rm d} \phi  \Pi_{j = 1} ^N \lambda\left( \phi, x_0 ^{-j} , x_0 ^{j} \right) e^{ - i \alpha \phi}. 
\label{eq04}
\end{equation}
We average Eq. (\ref{eq04}) with respect to the initial conditions
$x_0$ which are assumed independent identically distributed random
variable and we find
\begin{equation}
\langle P_N \left( \alpha \right) \rangle = { 1 \over 2 \pi} \int_{-\pi} ^\pi 
{\rm d} \phi \langle \lambda\left( \phi\right) \rangle^N e^{ - i \alpha \phi} 
\label{eq05}
\end{equation}
where from Eq. (\ref{eq03})
$$ \langle \lambda\left( \phi \right) \rangle = $$  
\begin{equation}
e^{ i \phi } \langle P_{LL} \rangle \langle P_{RL} \rangle + 
\langle P_{LL} \rangle \langle P_{RR} \rangle + \langle P_{LR} \rangle \langle P_{RL} \rangle + e^{ - i \phi} 
\langle P_{LR} \rangle 
\langle P_{RR} \rangle .
\label{eq06}
\end{equation}
Here $\langle P_{ij} \rangle$ is the probability of starting
in  $i=L,R$ (relative to the the Jepsen line) and
ending in $j=L,R$ 
and $\langle \cdots \rangle$ denotes an average over initial
condition. 
The probability $\langle P_{LR} \rangle$ is given in terms
of the Green function of the non interacting particle $g(x,x_0,t)$ 
and the initial
density of particles situated initially to the left of the Jepsen line 
$f_L(x_0)$
\begin{equation}
 \langle P_{LR} (vt) \rangle = \int_{-\overline{L}}  ^0 f_L (x_0) \int_{vt} ^{\overline{L}} g(x,x_0, t ) {\rm d} x {\rm d} x_0 .
\label{eq06g} 
\end{equation}
We see that
for  $\langle P_{LR} \rangle$ we average over initial conditions in the
domain $(-\overline{L},0)$ weighted by $f_L(x_0)$
 (since the starting point is to the left of the
Jepsen line) and also integrate over $x$ in the domain
$(vt, \overline{L})$ with the weight $g(x,x_0,t)$ 
(since the end point is to the right of the line). 
Similarly
\begin{equation} 
\langle P_{RR} (vt) \rangle = \int_0 ^{\overline{L}} f_{R} (x_0) \int_{vt} ^{\overline{L}} g(x,x_0 , t ) {\rm d} x {\rm d} x_0.
\label{eq07}
\end{equation}
As mentioned $f_L(x_0)$ $[f_R(x_0)]$
 is the PDF of initial positions of particles that initially
are at $x_0<0$ $(x_0>0)$ respectively,
hence $f_L(x_0) = 0$ when $x_0>0$ 
and $\int_{-\overline{L}} ^0 f_L (x_0) {\rm d} x_0 = 1$,
while  $f_R(x_0)$ is normalized and  non zero only in $(0,\overline{L})$.  
$\langle P_{LL} \rangle$ and $\langle P_{RR} \rangle$ are defined similarly,
and they satisfy 
$\langle P_{LL} \rangle = 1 - \langle P_{LR} \rangle$,
and $\langle P_{RL} \rangle = 1 - \langle P_{RR} \rangle$.

\section{Dynamics of the Tagged Particle} 
\label{SecRes} 

We now
apply the central limit theorem  to analyze 
the sum Eq. (\ref{eq01a}) using
Eq. (\ref{eq05}) 
when $N \to \infty$. 
The small $\phi$ expansion of the structure function
Eq. (\ref{eq06}) 
is
\begin{equation}
\langle \lambda(\phi) \rangle = 1 
+ i \langle \Delta \alpha \rangle \phi - {1 \over 2}\langle (\Delta \alpha)^2 \rangle \phi^2 + O(\phi^3)
\label{eq08}
\end{equation}
where 
\begin{equation}
\langle \Delta \alpha \rangle  = \langle P_{RL} \rangle - \langle P_{LR} \rangle
\label{eq09} 
\end{equation} 
as expected, 
and the variance 
$\sigma_{\Delta \alpha} ^2 =  \langle (\Delta \alpha)^2 \rangle- \langle \Delta \alpha \rangle^2 $ is
\begin{equation}
\sigma_{\Delta \alpha}  ^2 = \langle P_{RL}\rangle \langle P_{RR} \rangle +
 \langle P_{LR}\rangle \langle P_{LL} \rangle.
\label{eq10} 
\end{equation}
Using the central limit theorem 
we have the probability of zero total crossing of
the Jepsen line, namely $\alpha = 0$ in the limit $N \to \infty$
\begin{equation}
 P_N (\alpha = 0 ) \sim  
{1 \over \sqrt{ 2 \pi N} \sigma_{\Delta \alpha} }
 \exp\left( - { N \langle \Delta \alpha \rangle^2 \over 2 \sigma_{\Delta \alpha} ^2 } \right).
\label{eq11}
\end{equation}  
This simple result, valid for a large class of Green functions
and initial conditions, is  suited for the investigation
of a large number of single file problems. 

 The probability $P_N(\alpha =0)$ is according to our earlier 
discussion equal to the probability that the tagged particle
is in the interval $v t < x_T < vt + \delta x$, where $\delta x$ is
the distance between the Jepsen line and the position of the
second particle to the right of the Jepsen line. We assume that
$\delta x$ is small relative to the typical length scale
the tagged particle traveled. For example if the 
 density of interacting particles
$\rho$ is a constant, we assume $1/\rho << \sqrt{ \langle (x_T)^2 \rangle }$. 
 Hence we may replace $vt \to x_T$ to transform
$P_N(\alpha=0)$ to the PDF of the tagged particle
$P(x_T)$ \cite{remark}.

 Making the replacement $v t \to x_T$, using Eq. 
(\ref{eq11}) we have in the limit of large $N$ 
$$ P(x_T) \sim $$
\begin{equation}
C \exp\left\{ - { N \left[ \langle P_{LR} (x_T) \rangle - \langle P_{RL} (x_T) \rangle \right]^2 \over 2 \left[ \langle P_{LL} (x_T) \rangle \langle P_{LR} (x_T) \rangle + \langle P_{RR} (x_T) \rangle \langle P_{RL} (x_T) \rangle \right] } \right\}
\label{eq12}
\end{equation}
where $C$ is a normalization constant. In Eq. (\ref{eq12}) 
and what follows $\langle P_{ij} (x_T) \rangle$ is given by
Eqs.
(\ref{eq06g},\ref{eq07}) with $vt \to x_T$. 
Eq. (\ref{eq12}) yields
the PDF of the center particle $P(x_T)$ in terms of
 $\langle P_{ij} (x_T) \rangle$ which according to Eq. (\ref{eq07})
depends on the free
particle green function and the initial conditions. 
Thus the information contained in the non-interacting Green function
is sufficient for the determination of the single file diffusion
of the tagged particle.

\subsection{Thermal Equilibrium} 

 In the long time limit, and in the presence of a binding potential field,
e.g. harmonic field or particles in a box, an equilibrium is reached.
 Then initial conditions do not play a role. For example
$\langle P_{LR} (x_T) \rangle = P^{eq}_{R} (x_T)$ 
with 
\begin{equation}
P^{\rm eq}_R (x_T) = {1 \over Z} \int_{x_T} ^{\overline{L}} 
\exp\left( - { V(x) \over k_b T } \right) {\rm d} x 
\label{eq13a}
\end{equation} 
where $Z$ is the normalizing partition function 
\begin{equation}
Z= \int_{-\overline{L} } ^{\overline{L} } \exp\left[ - {V\left(x\right) \over k_b T }\right] {\rm d} x.
\label{eq13b}
\end{equation}
Similarly
\begin{equation}
P^{\rm eq}_L (x_T) = {1 \over Z} \int_{-\overline{L}} ^{x_T}
\exp\left( - { V(x) \over k_b T } \right) {\rm d} x.
\label{eq14a}
\end{equation} 
In Eqs. (\ref{eq13a},\ref{eq14a}) we used the steady state solution,
 $\lim_{t \to \infty}  g(x,x_0,t) = \exp[ - V(x)/k_b T] / Z$, which is
Boltzmann's distribution suited for a system in thermal equilibrium.
Using Eq. (\ref{eq12})
the position PDF of the tagged  particle is
$\lim_{t \to \infty} P(x_T) = P^{\rm eq}(x_T)$ 
\begin{equation}
P^{\rm eq} (x_T) \sim  C \exp\left\{ - { N \left[ \langle P_{R}^{eq}  (x_T) \rangle - \langle P_{L} ^{eq}  (x_T) \rangle \right]^2 \over 4  \langle P_{L} ^{eq} (x_T) \rangle \langle P_{R} ^{eq}  (x_T) \rangle } \right\}.
\label{eq15a}
\end{equation} 
If the potential is symmetric $V(x) = V(-x)$, e.g. particles in a box or
harmonic field, we have for 
not too large $x_T$, $P_{L} ^{eq} (x_T) \simeq 1/2$, 
$P_{R} ^{eq} (x_T) \simeq 1/2$ hence
from Eq. (\ref{eq12})
\begin{equation}
P^{\rm eq} (x_T) \sim C \exp \left\{ - N \left[ \langle P_R ^{eq} (x_T) \rangle -
\langle P_L (x_T) \rangle \right]^2 \right\}.
\label{eq16a}
\end{equation}
Expanding the expression in the exponent in $x_T$ (since $N$ is large) we find
using Eqs. (\ref{eq13a},\ref{eq14a},\ref{eq15a}) 
\begin{equation}
P^{\rm eq} ( x_T ) \sim {2 \sqrt{N} \over \sqrt{\pi} Z} \exp \left[ - { 4 N \over Z^2} \left( x_T \right)^2 \right]
\label{eq17a}
\end{equation} 
where with out loss of generality we assigned $V(x=0)=0$. 
Hence the standard deviation is 
\begin{equation}
\langle (x_T)^2 \rangle \sim {Z^2 \over 8 N} .
\label{eq17aa}
\end{equation} 
The same expression is found in Appendix A using the many body
Boltzmann distribution, and integrating over all the particles
except the tagged particle. 

\subsection{Simple Illustration}
\label{secSI}

 We now consider the situation of particles free of a force $F(x)=0$ with open
boundary conditions $\overline{L} \to \infty$ where initially all the particles
are on the vicinity of  the origin. 
More precisely, the tagged particle is initially situated at
$x_T =0$, $N$ particles to its right on $\epsilon\to 0^{+}$ and
$N$ particles on $-\epsilon$. This problem was solved already by Aslangul \cite{Aslangul},
and here we recover the known result using our formulas. 
The Green function $g(x,x_0,t)$ of
a free particle is
\begin{equation}
g(x,x_0,t) = {\exp \left[ - { \left( x - x_0\right)^2 \over 4 D t} \right] \over \sqrt{ 4\pi D t} } 
\label{eqAs01}
\end{equation}
where  as mentioned $D$ is the diffusion coefficient of the free particle. 
With the specified initial conditions we
have
$$ \langle P_{RL} \left( x_T \right) \rangle = \lim_{\epsilon \to 0} \int_0 ^\infty \delta(x_0 - \epsilon) \int_{-\infty} ^{x_T} { \exp\left[ - { \left( x - x_0 \right)^2 \over 4 D t }  \right] \over \sqrt{4\pi D t} } {\rm d} x {\rm d} x_0 = $$
\begin{equation} 
{1 \over 2} + \int_0 ^{x_T} { e^{ - { x^2 \over 4 D t}} \over \sqrt{ 4 \pi D t } } {\rm d} x. 
\label{eqAs02}
\end{equation}
Similarly
\begin{equation} 
 \langle P_{LR} \left( x_t \right) \rangle = 
{1 \over 2} - \int_0 ^{x_T} { e^{ - { x^2 \over 4 D t}} \over \sqrt{ 4 \pi D t } } {\rm d} x. 
\label{eqAs03}
\end{equation}
When $x_T\ll \sqrt{2 D t}$ we Taylor expand in $x_T$ to find $(\langle P_{LR} \rangle - \langle P_{RL} \rangle)^2 \sim
(x_T)^2/ \pi D t $, using Eq. (\ref{eq12}) 
we recover the result in \cite{Aslangul}
\begin{equation}
P(x_T) \sim { 1 \over \pi} \sqrt{ {N \over D t}} \exp\left[ - { N (x_T)^2 \over \pi D t } \right],
\label{eqAs03}
\end{equation}
hence 
\begin{equation}
\langle (x_T)^2 \rangle \sim  {\pi D t \over 2 N }.
\label{eqAs03a} 
\end{equation}
The diffusion is
normal, in the sense that the mean square displacement increases linearly
in time. However the diffusion of the tagged particle
is slowed down compared
with a free particle, by a factor of $1/N$ which is  due to the collisions
with all other Brownian particles in the  system. Clearly the approximation
breaks down if one is interested in the tails of $P(x_T)$, since we
used $x_T \ll \sqrt{ 2 D t}$. Though clearly
 when $N$ is large,
the probability of finding such a particle is extremely small (i.e.
use Eq. (\ref{eqAs03}) $P(x_T = \sqrt{ 2 D t}) \sim \exp( -N 2/\pi)$).

\subsection{Formula for $\langle (x_T)^2 \rangle$ }

 We now consider
 symmetric potential fields $V(x)=V(-x)$, and symmetric initial 
conditions.
 The latter means that the density of
particles at time $t=0$ to the left of the tagged particle, i.e. those
residing in $x_0<0$, is the same as for those residing to the right, 
$f_R(x_0) = f_L ( - x_0)$ (e.g. uniform initial conditions). In this
case the subscript $R$ and $L$ is redundant and we use $f(x_0)=f_R(x_0)=f_L(-x_0)$,
where $f(x_0)=0$ if $x_0<0$, $f(x_0)\ge 0$ and $\int_0 ^\infty f(x_0) {\rm d} x_0=1$.
From symmetry it is clear that the tagged particle is unbiased,
namely $\langle x_T \rangle = 0 $. Further, since $N$ is large we may
expand the expressions in the exponent in Eq. (\ref{eq12}) 
in $x_T$ to obtain
the leading term
$$ \langle P_{RL} \rangle - \langle P_{LR} \rangle = $$
\begin{equation}
{\partial \over \partial x_T} \left[ \langle P_{RL} \left( x_T \right) \rangle -
  \langle P_{LR} \left( x_T \right) \rangle\right]|_{x_{T}=0} 
  x_T + O[(x_T)^2]. 
\label{eq12a}
\end{equation}   
Similarly
for small $x_T$
\begin{equation}
\begin{array}{c}
 \langle P_{LL} \rangle \langle P_{LR} \rangle + \langle P_{RR} \rangle \langle P_{RL} \rangle= \\
2 \langle P_{RR} \left( x_T \right) \rangle \left[ 1 - \langle P_{RR}\left( x_T \right) \rangle \right] |_{x_T =0} + O(x_T) 
\end{array}
\label{eq14}
\end{equation}
To derive Eq. (\ref{eq12a}) we used the symmetry of the problem,
 which means that
the probability of crossing the point $x_T=0$ from left to right
is the same as  the probability to cross from right to left 
$\langle P_{LR} \rangle_{x_T =0} =\langle P_{RL} \rangle_{x_T = 0} $ and
similarly
$\langle P_{LL} \rangle_{x_T =0}  = 
\langle P_{RR} \rangle_{x_T =0}$ by symmetry. 
We designate the reflection coefficient,
\begin{equation}
{\cal R} = \langle P_{RR} \rangle_{x_T = 0}  
\label{eq15}
\end{equation}
[or ${\cal R} = \langle P_{LL} \rangle_{x_T = 0}$]  
since it is the probability that a particle starting at $x_0>0$, is found in $x>0$ at time
$t$ when an average over initial conditions is made
\begin{equation}
{\cal R} = \int_0 ^{\overline{L}} f(x_0) \int_0 ^{\overline{L}} g\left( x, x_0 , t \right) {\rm d} x {\rm d} x_0.
\label{eq16}
\end{equation}
A transmission coefficient 
is defined through 
${\cal T} = \langle P_{RL}  \rangle_{x_T = 0} = 
 \langle P_{LR} \rangle_{x_T = 0} $ 
which is related to the reflection coefficient in the usual way 
$ {\cal T} = 1 -{\cal R}$. 

Turning our attention to Eq. (\ref{eq12a}), from left-right symmetry we have
\begin{equation}
{\partial \over \partial x_T } \langle P_{RL} \left( x_T \right) \rangle|_{x_T = 0} 
= - {\partial \over \partial x_T } \langle P_{LR} \left( x_T \right) \rangle|_{x_T = 0}.
\label{eq18}
\end{equation}
Hence we define 
\begin{equation}
{\it r} = 
{\partial \over \partial x_T } \langle P_{RL} \left( x_T \right) \rangle|_{x_T = 0} 
\label{eqj}
\end{equation}
where from  Eq. 
(\ref{eq07})
\begin{equation}
{\it r}   = \int_0 ^{\overline{L}} f(x_0) g(0,x_0, t ) {\rm d} x_0 .
\label{eq19}
\end{equation}
So  ${\it r}$ is the density of non interacting particles at $x=0$ 
for an initial density $f(x_0)$. 
Note that 
since  
$\langle P_{RL}(x_T) \rangle + \langle P_{RR}(x_T) \rangle = 1$
we have  
\begin{equation}
{\it r} = - {\partial \over \partial x_T}  \langle P_{RR} (x_T ) \rangle|_{x_T = 0}.
\label{eq19u}
\end{equation}

Inserting Eq.
(\ref{eqj})
in  Eq. (\ref{eq12a}) using Eq. 
(\ref{eq18}) 
we have $\langle P_{RL} \rangle - \langle P_{LR} \rangle = 2 r x_T + \cdots$.
Inserting Eq. (\ref{eq15}) in Eq. (\ref{eq14}) 
we find our main result: the 
probability density function  of the position of the tagged central particle
\begin{equation}
P(x_T) \sim { 1 \over \sqrt{ 2 \pi \langle (x_T)^2 \rangle }}
\exp\left[ - {\left( x_T \right)^2 \over 2 \langle \left( x_T \right)^2 \rangle} \right],
\label{eq20}
\end{equation}
where
\begin{equation}
\langle (x_T)^2 \rangle = { {\cal R} \left( 1 - {\cal R} \right)\over 2 N {\it r} ^2 } 
\label{eq21}
\end{equation}
is the  mean square displacement of the tagged  particle. 
 The single particle probability $\langle P_{RR}(x_T) \rangle$ 
 gives ${\cal R}$ Eq. 
(\ref{eq15})
and ${\it r}$ Eq.
(\ref{eq19u}) which in turn yield the mean square displacement of
the tagged particle Eq. (\ref{eq21}). 
We will soon use this equation to demonstrate a variety of physical behaviors. 
However first we establish a connection between our work
and the theory of order statistics. 

\subsection{Order Statistics}
\label{SecGumbel}

 Generate $n$ independent identically distributed random variables
drawn from a PDF $\hat{r}(x)$ and arrange them in
increasing order.  Order statistics deals with
the m-th observable $m=1,\cdots,n$ among $n$ observations taken
in the increasing order, which is denoted $x_m$. The PDF of $x_m$, $\phi(x_m)$ 
depends on the PDF $\hat{r}(x)$, the sample size $n$ and
the order $m$. Let $\hat{R}(x)$ be the cumulative distribution
of $x$, e.g. if the domain of $x$ is $-\infty<x<\infty$, $\hat{R}(x)=
\int_{-\infty} ^x \hat{r}(x) {\rm d} x$ as usual. 
Following well known result \cite{Gumbel}, define $\hat{x}$ 
with 
\begin{equation}
\hat{R}(\hat{x} ) = { m\over n+1}
\label{eqOrder1} 
\end{equation}
then when $n$ is large and $m/n $ is of the order $1/2$
\begin{equation} 
\phi(x_m) = \mbox{const} \exp\left\{ - { n \left( x- \hat{x} \right)^2 \hat{r}^2 (\hat{x} ) \over 2 \hat{R}(\hat{x} ) \left[ 1 - \hat{R}(\hat{x} ) \right] } \right\}.
\label{eqorder2} 
\end{equation} 
Hence the variance of $x_m$ 
\begin{equation}
(\sigma_m)^2  = { \hat{R}(\hat{x}) \left[ 1 - \hat{R}\left( \hat{x} \right) \right] \over n \hat{r}^2 \left( \hat{x} \right) }
\label{eqorder3} 
\end{equation} 
which has some resemblance to Eq. (\ref{eq21}).

 The problem of the motion of a tagged particle is mathematically
identical to the problem of order statistics in two cases: (i)
in the presence of a binding potential and in the long
time limit and (ii) when all the particles start on the
same point. In both cases the single particle
 PDF $g(x,x_0,t)$ of all
the particles are identical since it is either  independent of $x_0$ (case (i)) 
or we have a unique initial condition (case treated in  
sub section \ref{secSI}). 
For example in the presence
of a binding field $\lim_{t \to \infty} g(x,x_0,t) = \exp[- V(x)/k_b T ] / Z$
which is independent of the initial position of the particle.
Hence in equilibrium, to find the center particle we may draw $(2 N+1)$
random variables from Boltzmann's distribution, and search for the center
particle (which will give the position of the interacting tagged particle). 
Or, using the language of order statistics we have  $\hat{r}(x) = 
\exp[- V(x)/k_b T ] / Z$.
Using  a symmetric potential $V(x)=V(-x)$  
and Eq. 
(\ref{eqOrder1})  
we insert  $n=2 N+1$, and $m=N+1$; hence, when $N$ is large
we have
\begin{equation}
\hat{R}(\hat{x} ) = \int_{-\infty} ^{{\hat x}} { \exp\left[ - {V(\hat{x})\over  k_b T} \right] \over Z} {\rm d} x  = {1\over 2} .
\label{eqorder4} 
\end{equation}
Thus, $\hat{x} = 0$. Then using Eq. 
(\ref{eqorder3}) we find
\begin{equation} 
\langle (x_T)^2 \rangle = {Z^2 \over 8 N} 
\end{equation} 
which is the same as Eq. 
(\ref{eq17aa}) (recall $V(0) =0$). 

While  Eq. (\ref{eqorder3}) has superficially a structure similar to Eq. 
(\ref{eq21}) they are different. Our ${\cal R}$ 
Eq. 
(\ref{eq15})
is generally not equal half, neither
must it be close to that value 
[so Eq. (\ref{eqOrder1}) is not generally related to our problem].  
In fact when $t\to 0$ we must have $\langle (x_T)^2 \rangle \to 0$ which
is found when $\lim_{t \to 0} {\cal R} = 1$ (see examples below). 
In the problem of motion of a tagged particle, the number of particles
which can interact with the center particle is usually increasing
with time (hence $n$ is not fixed). For example consider the classical
case of  uniformly
distributed particles in infinite space and in the absence of forces.
Roughly speaking  particles  at distances of the order of 
$l_{{\rm eff}} = \sqrt{ Dt}$ or shorter 
can influence the tagged particle motion via collisions.
So in this  case roughly $ N_{{\rm eff}}=\rho \sqrt{D t}$ 
particles participate in the
process. In contrast, if all particles are initially at the vicinity 
of the origin $N$ particles participate i.e. influence
the motion of the tagged particle. Interestingly
this gives an argument for the well known behavior
of the mean square displacement of the tagged particle, with uniform density of
particles,  namely use Eq. 
(\ref{eqAs03a})  $\langle (x_T)^2 \rangle \sim \pi D t / 2 N$ 
and replace
$N$ with $N_{{\rm eff}} = \rho \sqrt{D t}$ 
to find $\langle (x_T)^2 \rangle \simeq    
\sqrt{ D t} / \rho$ 
which of course misses the correct numerical prefactor (see Eq. 
\ref{eqbox06}
below). 

\section{Physical Illustrations}
\label{secPI}

\subsection{Particles in a Box}

 Consider particles in a finite box extending from $-\overline{L}$ to
$\overline{L}$ which was recently treated with the Bethe ansatz and
numerical simulations by Lizana  and Ambj\"ornson  \cite{LIAM}. 
The tagged particle initially
at $x_T =0$ has $N$ particles to its right and $N$ to its left.
These particles are assumed uniformly distributed, hence 
\begin{equation}
f(x_0)= \left\{
\begin{array}{l l}
{1\over \overline{L}} \  & \ 
0<x_0 < \overline{L} \\
\ & \ \\
0 \ & \ \mbox{otherwise} .
\end{array}
\right.
\label{eqbox0}
\end{equation}
In the limit $\overline{L} \to \infty$
and $N\to \infty$ in such a way that the density $\rho= N / \overline{L}$ is fixed,
we obtain single file diffusion in an infinite system,
a case well studied long ago \cite{Ha,Le}. 

 The single particle Green function of a particle in a box,
with reflecting boundary conditions 
$\partial g(x,x_0,t) / \partial x|_{x=\pm
\overline{L}}= 0$ is solved using an eigenfunction expansion 
\cite{Risken}
$$ g(x,x_0,t) = {1 \over 2 \overline{L} } + $$
\begin{equation}
{1 \over \overline{L} } 
\sum_{n=1} ^\infty \cos \left[ { n \pi \over 2\overline{L}} \left( x + \overline{L} \right) \right] \cos\left[ { n \pi \over 2 \overline{L} } \left( x_0 + \overline{L} \right) \right] \exp\left( - D { n^2 \pi^2 \over 4 \overline{L}^2 } t \right).
\label{eqbox01}
\end{equation}
With Eqs. (\ref{eq16},\ref{eqbox0},\ref{eqbox01}) we find
\begin{equation}
{\cal R}(t) = {1 \over 2} + { 4 \over \pi^2} \sum_{n=1, \mbox{Odd} } ^\infty
{ \exp\left( - D { n^2 \pi^2 \over 4 \overline{L}^2 } t \right) \over n^2}
\label{eqbox02}
\end{equation}
where the summation is over odd $n$. 
At $t=0$, ${\cal R} =1$ since all particles
initially in $(0, \overline{L})$ did not have
time to move to the other side of the box, and
$\lim_{t \to \infty} {\cal R} = 1/2$ since in the
long time limit there is equal probability for
a non interacting particle to occupy each half of
the box. 
Using Eqs. (\ref{eq19},\ref{eqbox0},\ref{eqbox01}) we find 
\begin{equation}
r= {1 \over  2 \overline{L} } 
\label{eqbox02aa}
\end{equation}
hence Eq. (\ref{eq21}) gives
the mean square displacement of the tagged particle
\begin{equation}
\langle \left( x_T \right)^2 \rangle \sim 2 {{\cal R}(t) \left[ 1 - {\cal R}(t) \right] \overline{L}^2 \over N}.
\label{eqbox03}
\end{equation}
The eigenvalues of the non-interacting particle determine the
multi exponential type of  decay of 
${\cal R}(t)$ with time,  which in turn determines the
dynamics of the interacting tagged particle \cite{remark3}.   
\begin{figure}
\begin{center}
\epsfig{figure=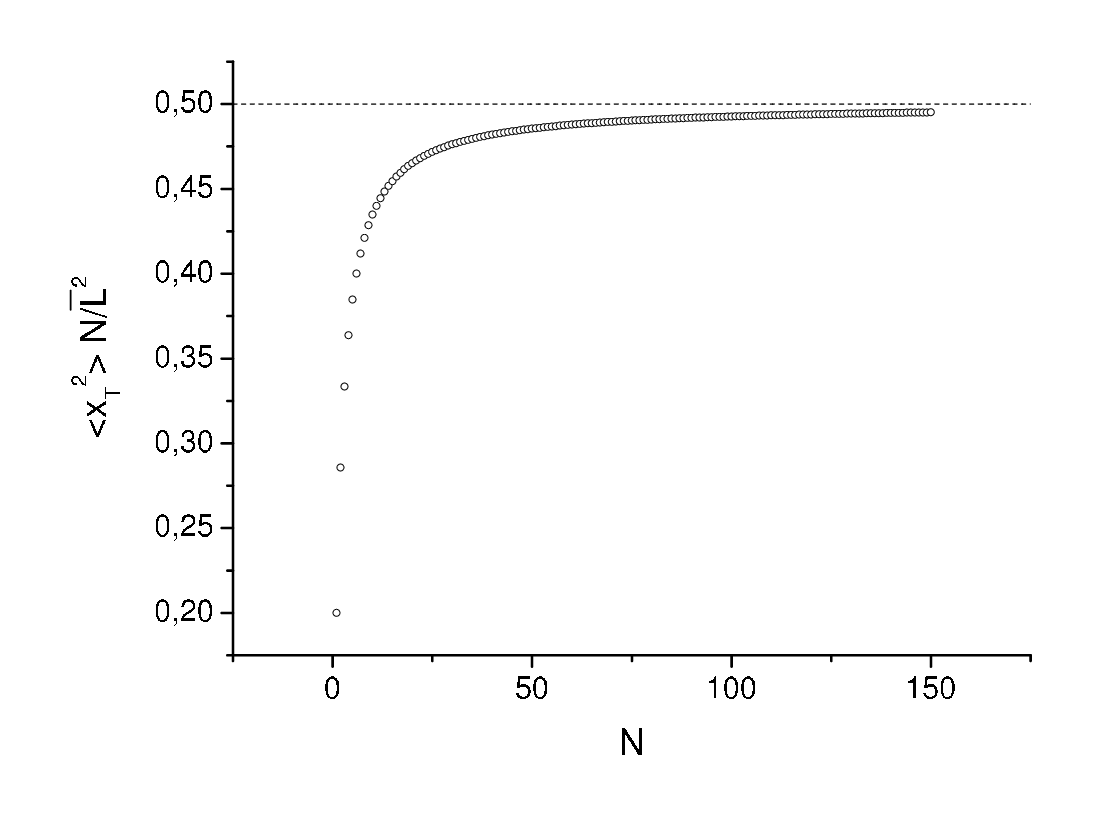,totalheight=0.34\textheight, width=0.45\textwidth}
\end{center}
\caption{ For particles in a box  
we show $\langle (x_T)^2 \rangle N / \overline{L}^2$ of the tagged particle, in equilibrium versus $N$. Exact expression
obtained in \cite{LIAM} valid for all $N$ (dots) converges in the limit
of $N \to \infty$ to the behavior predicted by our theory (dashed line) 
Eq. (\ref{eqbox07a}). For $N=70$ clear deviations between
our asymptotic theory and the exact result in \cite{LIAM}
are found. 
}
\label{fig11}
\end{figure}

 Let $\delta^2 = D \pi^2 t / 4 \overline{L}^2$ hence the limit $\delta \ll 1$ gives
the short time dynamics before the particles interact with the walls,
or equivalently the limit of an infinite system. 
The reflection coefficient is rewritten
\begin{equation}
{\cal R} = { 1 \over 2} + { 4 \over \pi^2} \left( 
\delta \sum_{n=1, \mbox{Odd}} ^\infty { e^{ - \delta^2 n^2} - 1 \over \delta^2 n^2 } \delta + \sum_{n=1, Odd} ^\infty { 1 \over n^2} \right).
\label{eqbox04}
\end{equation}
When $\delta\ll 1$ we may replace the first summation with integration
\begin{equation}
\sum_{n=1,\mbox{Odd}} ^\infty { e^{- \delta^2 n^2 } - 1 \over \delta^2 n^2} \delta \simeq {1 \over 2} \int_0 ^\infty { e^{- y^2} - 1 \over y^2} {\rm d} y = { - \sqrt{\pi} \over 2} 
\label{eqbox05}
\end{equation}
where the factor $1/2$ on the RHS  comes from  the summation over  only odd $n$ on the LHS. 
Using $\sum_{n=1,\mbox{Odd} } ^\infty 1/n^2 = \pi^2/ 8$ we have
${\cal R} = 1 - { \sqrt{D t} \over \overline{L}  \sqrt{\pi} }+ \cdots$,
and hence 
\begin{equation}
\langle \left( x_T \right)^2 \rangle \sim { 2 \over \sqrt{\pi} } 
{\sqrt{ D t} \over \rho} .
\label{eqbox06}
\end{equation} 
This result was obtained in \cite{Ha,Le}, and 
it describes the dynamics of the tagged particle in a box,
before particles have time to interact with the walls,
namely when $\delta$ is small.

 In the opposite limit of long times and finite systems
we attain equilibrium, then $\lim_{t \to \infty} {\cal R} = 1/2$
and
\begin{equation}
P(x_T) \sim { \sqrt{N} \over \sqrt{\pi} \overline{L} } e^{ - N \left( x_T\right)^2 / \overline{L}^2 }, 
\label{eqbox07}
\end{equation}
hence 
\begin{equation}
\lim _{t \to \infty} \langle (x_T)^2 \rangle = \overline{L}^2 / 2 N.
\label{eqbox07a}
\end{equation}
Eq. (\ref{eqbox07}) is a special case of the more general Eq.
(\ref{eq17a}) since the single particle normalizing
 partition function
for our example is $Z= 2 \overline{L}$. 

 As mentioned, in \cite{LIAM} the Bethe ansatz was used to 
solve the problem of tagged particle motion in a box.
Among other things an
 exact expression for the long time limit of $\langle (x_T)^2 \rangle$,
 valid for all
$N$ was found \cite{LIAM} 
\begin{equation}
\lim _{t \to \infty} \langle (x_T)^2 \rangle = \overline{L}^2 \left( {1 \over 4}\right)^{N+1}  { \Gamma\left({1\over 2} \right) \Gamma\left[ 2\left( N + 1 \right) \right]  \over \Gamma\left( N + 1 \right) \Gamma\left( N + {5 \over 2} \right) }. 
\label{eqbox07liz}
\end{equation}
Since our formalism is valid only in the large $N$ limit,
comparison of our solution to an exact result like Eq.
 (\ref{eqbox07liz}) provides
insight to the question of convergence. 
In Fig. \ref{fig11} we plot 
$   \lim_{t \to \infty } \langle (x_T)^2  \rangle N / \overline{L}^2$  
versus $N$ using Eq. (\ref{eqbox07liz}) comparing it to our
Eq. (\ref{eqbox07a}) which gives
$  \lim_{t \to \infty } \langle (x_T)^2 \rangle N/ \overline{L}^2  =1/2 $. 
Not surprisingly we see that the two results yield asymptotically
the same result. The figure illustrates
that even for $N=100$ deviations between exact results and the present
theory are observable.

\begin{figure}
\begin{center}
\epsfig{figure=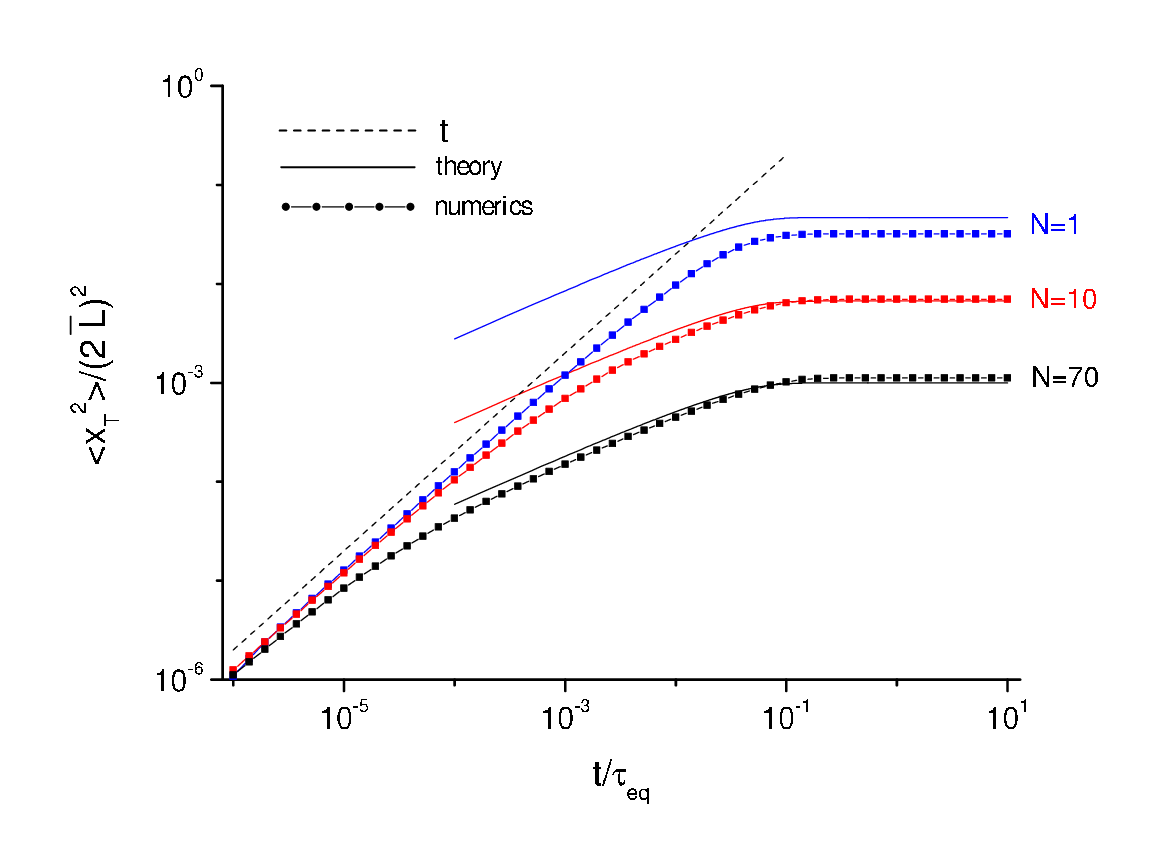,totalheight=0.34\textheight, width=0.45\textwidth}
\end{center}
\caption{ Motion of tagged particle in a box is shown for $N=1,10,70$.
Simulation results (squares) are taken from \cite{LIAM} (see text). At
short
time simulations  exhibit normal diffusion $\langle (x_T)^2 \rangle = 2 D t$
 as illustrated by the straight dashed line which is a guide to the eye. 
Our theory is expected to work well in the large $N$ limit 
and when many collision events between tagged particle and surrounding
Brownian particles took place.  
Hence agreement between theory (solid line) and simulation
is reasonable at most only for $N=70$ and not for too short times.  
} 
\label{fig10}
\end{figure}

 In \cite{LIAM} numerical simulation of systems with  $N=1,10,70$
(maximum  of  $141$
particles) were performed, and favorably
compared with the Bethe ansatz solution.
These simulations were made for finite size hard core particles
whose diameter is $\Delta$. To make comparison between our theory
and simulation we scale  the size of the box according to $2 \overline{L} \to
2 \overline{L} - (2 N ) \Delta$. In Fig. \ref{fig10}  the scaled
mean square displacement versus $t/\tau_{eq}$ is shown, where $\tau_{eq} = 4 \overline{L}^2 / D$.
 Some general features
of our theory can now be discussed. First, at short
times simulations show normal behavior $\langle (x_T)^2 \rangle = 2 D t$,
which is a trivial effect: particles did not have time to
collide and hence the tagged particle diffuses normally as
if it is free. After particles start to collide, deviations
from normal diffusion are observed 
(for $N=70$ but  as expected not for $N=1$). These 
are mainly due to collisions with other particles. Finally
saturation due to the finite size of the system is found. While
our theory shows the general trend of simulations (say for $N=70$
and for not too short times) it is clearly
not in perfect agreement. We argue that this is due to
the small number of particles $N=70$,
since as we showed in Fig. \ref{fig11}, at least for large
times and large $N$ our theory gives the exact result.  
It would be nice to have simulations with point particles,
with larger $N$ and since there are three phases of the motion:
normal diffusion, single file diffusion without the  influence of 
the walls  
and finally saturation to equilibrium. 
Separation of time scales is needed to demonstrate
these behaviors clearly.

\subsection{Gaussian Packet}

 Consider particles without  external forces $V(x)=0$, in an infinite system.
Initially particles are spread with a Gaussian packet with width $\xi$:
\begin{equation}
f(x_0) = { \sqrt{2} \over \sqrt{\pi} \xi } \exp\left[ - { (x_0) ^2 \over 2 \xi^2} \right] 
\label{eqGP1} 
\end{equation}
for $x_0 >0$. As before we consider the tagged particle motion,
which is initially at $x_T=0$, with $N$ particles to its
left and $N$ to its right. With the free particle Green function 
$g(x,x_0,t)$ Eq. (\ref{eqAs01}) we proceed to find $\langle (x_T )^2 \rangle$.

\begin{figure}
\begin{center}
\epsfig{figure=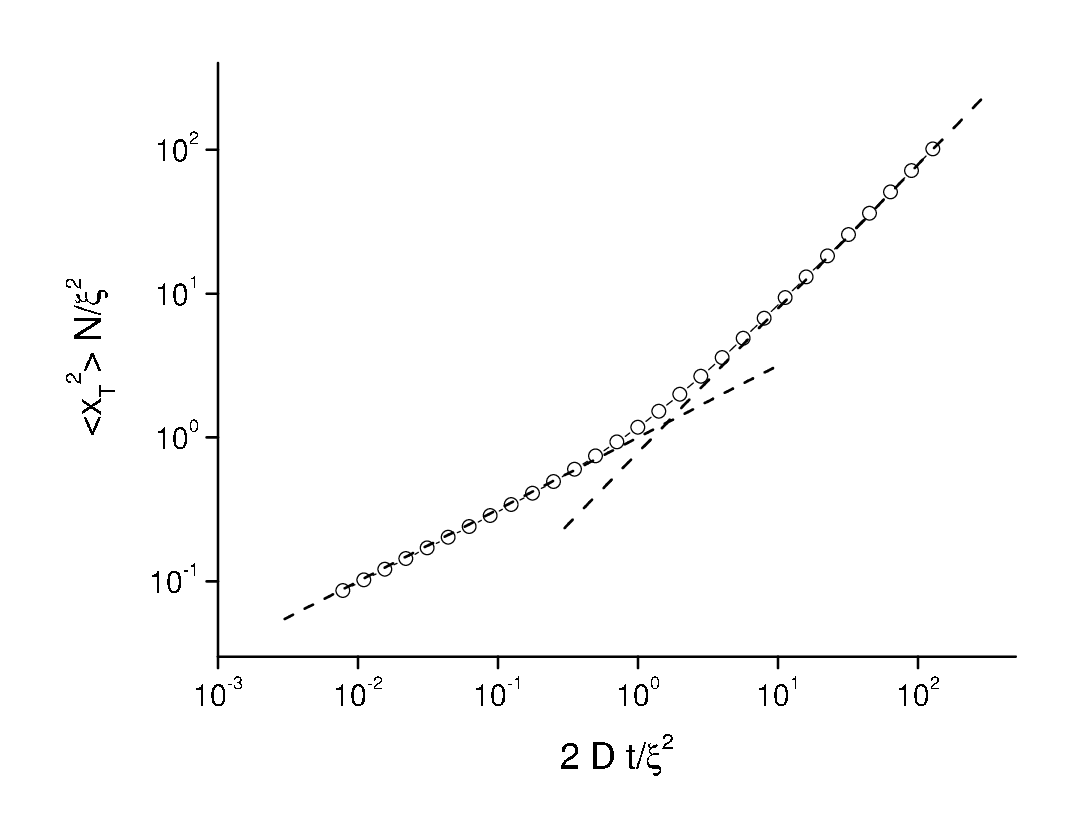,totalheight=0.34\textheight, width=0.45\textwidth}
\end{center}
\caption{ Scaled mean square displacement of the tagged particle
with Gaussian initial
conditions of the packet of particles exhibits a transition
between 
short time $\langle (x_T)^2 \rangle\propto
t^{1/2}$ law to $\langle (x_T)^2 \rangle \propto t$ behavior. 
Dashed lines are short and long time asymptotic behavior Eq.  
(\ref{eqGPii}),
circles represent Eq. (\ref{eqGP6}). 
} 
\label{fig12a}
\end{figure}

  The reflection probability Eq. (\ref{eq16}) is
\begin{equation}
{\cal R} = \int_0 ^\infty {\rm d} x_0 \sqrt{{2 \over \xi^2  \pi}} e^{ - {(x_0)^2 \over 2 \xi^2 } } 
\int_0 ^\infty { {\rm d} x \over \sqrt{ 4 \pi D t } } e^{ - { (x- x_0)^2 \over 4 D t} }  .
\label{eqGP2}
\end{equation}
Changing variables  according to $y^2/2 = (x-x_0)^2/ 4 Dt$
and using dimensionless parameter $\tilde{\xi} = \xi/ \sqrt{ 2 D t}$, we find
\begin{equation}
{\cal R} = {1 \over 2} + { 1 \over \tilde{\xi}} \sqrt{ { 2 \over \pi}} \int_0 ^\infty
{\rm d} \tilde{x}_0 e^{ - { (\tilde{x}_0)^2 \over 2 \tilde{\xi}^2 }} { \mbox{Erf}\left( \tilde{x}_0/\sqrt{2} \right) 
\over 2} 
\label{eqGP3}
\end{equation}
where 
$\mbox{Erf}(\tilde{x}_0/\sqrt{2} )/2=\int_0 ^{\tilde{x}_0 } e^{ - y^2 /2 } {\rm d} y/\sqrt{ 2 \pi} $
is the error function \cite{Abr}.
Mathematica solves the integral
in Eq. (\ref{eqGP3}) and we find
\begin{equation}
{\cal R} = {1 \over 2} + {1 \over \pi} \mbox{arccot}\left( { \sqrt{ 2 D t} \over \xi } \right).
\label{eqGP4}
\end{equation} 
For short times $\sqrt{D t}\ll \xi$ we have ${\cal R}\sim 1 - { \sqrt{ 2 D t} \over \pi \xi } $,  
namely most particles did not have time to cross the origin, while in the
opposite limit $\lim_{t \to \infty} {\cal R} = 1/2$ due to the symmetry of
initial conditions. 
The calculation of $r$ Eq. (\ref{eq19}) using Eqs. (\ref{eqAs01},\ref{eqGP1}) is straightforward
\begin{equation}
r = { 1  \over 2 \sqrt{ \pi D t} \sqrt{ 1 + \xi^2 / 2 D t}  },
\label{eqGP5}
\end{equation} 
Inserting Eqs. (\ref{eqGP4},\ref{eqGP5}) in Eq. 
(\ref{eq21}) we find
\begin{equation} 
\langle ( x_T )^2 \rangle \sim \xi^2 {\pi \over N} \left( 1 + { 2 D t \over \xi^2} \right) \left[ {1 \over 4} -
{1 \over \pi^2 } \mbox{arccot}^2 \left( \sqrt{{ 2 D t \over \xi^2} } \right) \right], 
\label{eqGP6}
\end{equation} 
This solution is shown in Fig. \ref{fig12a}
with its two  limiting behaviors
\begin{equation}
\langle (x_T)^2 \rangle \sim \left\{ 
\begin{array}{l l}
\xi { \sqrt{ 2 D t} \over N} & \mbox{short times} \ \  2 D t \ll \xi^2 \\
{\pi D  \over 2 N} t & \mbox{long times} \ \  2 D t \gg \xi^2 .
\end{array}
\right.
\label{eqGPii}
\end{equation}
For short times the particles do not have time to disperse;
hence, the motion of the tagged particle is slower than normal, increasing
as $t^{1/2}$ which is similar to the single file diffusion
with a uniform density Eq.  
(\ref{eqbox06}). Roughly speaking,
for short times 
the tagged particle sees a uniform density of particles with $\rho=N/\xi$.
For long times we recover the behavior in Eq. 
(\ref{eqAs03a}) since the scale of diffusion is much larger than $\xi$.
Hence if we start with   a Gaussian or delta function packet
we get in the long time limit similar behavior, as we showed.  

\subsection{Particles in Harmonic Oscillator}

 Consider particles in a harmonic potential $V(x) = m \omega^2 x^2 / 2$ where
$\omega>0$ is the harmonic frequency. The single particle  undergoes
an Ornstein Uhlenbeck process \cite{Risken} and the corresponding
single particle Green
function is
$$ g(x,x_0, t) = $$
\begin{equation}
{ 1 \over \sqrt{ 2 \pi D \tau \left( 1 - e^{ - 2 t /\tau} \right)}} \exp\left[ - { \left( x - x_0 e^{ - t / \tau} \right)^2 \over 2 D \tau \left( 1 - e^{ - 2 t / \tau } \right) } \right],
\label{eqHO01}
\end{equation}
where $(\tau)^{-1} = D m \omega^2/ k_b T $ is the inverse relaxation time.

\begin{figure}
\begin{center}
\epsfig{figure=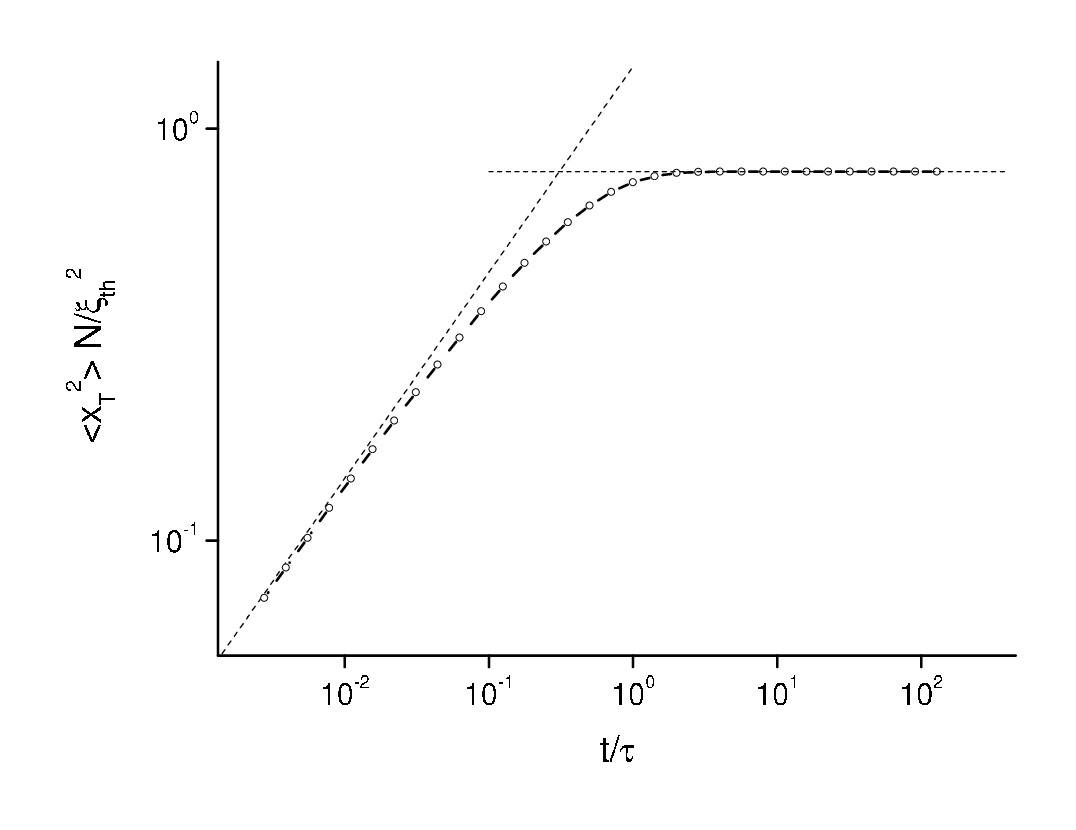,totalheight=0.34\textheight, width=0.45\textwidth}
\end{center}
\caption{ Scaled mean square displacement  of tagged particle in harmonic field
Eq. 
(\ref{eqHO06}) 
exhibits a transition between a short time $\langle (x_T)^2 \rangle\propto
t^{1/2}$ law to saturation due to the binding field. 
The horizontal dashed line is the long time $\langle (x_T)^2 \rangle = \pi \xi^{2}  _{th} /4 N$ behavior,  
the dashed line is the  short time limit  Eq. (\ref{eqHO07}),
and the dotted dashed line is Eq.  
(\ref{eqHO06}).
} 
\label{fig12}
\end{figure}

 We assume thermal initial conditions 
\begin{equation}
f(x_0) = { 2 \sqrt{ m \omega^2 } \over \sqrt{ 2 \pi k_b T} } \exp\left( - {m \omega^2 x^2 \over 2 k_b T} \right), \ \ \ x_0 > 0.
\label{eqHO02} 
\end{equation} 
Using Eq. (\ref{eq19}) it is easy to show that
\begin{equation}
r = { 1 \over \sqrt{ 2 \pi} \xi_{{\rm th}} }, 
\label{eqHO03} 
\end{equation} 
where the thermal length is $\xi_{{\rm th}} = \sqrt{ D \tau} = \sqrt{ m \omega^2 / k_b T} $.
 Note that one can write $r = 1/ Z$ which  as we will soon prove
is a general result
valid for all potential fields satisfying $V(x)=V(-x)$,
 provided that the initial condition
$f(x_0)$ is the thermal equilibrium . 
The reflection coefficient [Eq. 
(\ref{eq16})] is 
\begin{equation}
{\cal R} = {1 \over 2} + \sqrt{{1\over 2 \pi} } \eta \int_0 ^\infty e^{ - {\eta^2 y^2 \over 2} } \mbox{Erf} \left( { y \over \sqrt{2} } \right) {\rm d} y,
\label{eqHO04} 
\end{equation} 
where $\eta = e^{t/\tau} \sqrt{ 1 - e^{ -2 t / \tau}}$. Using
Mathematica 
\begin{equation}
{\cal R} = {1 \over 2} + { 1 \over \pi} \mbox{arccot} \left( \sqrt{ e^{ 2 t /\tau } - 1} \right). 
\label{eqHO05} 
\end{equation}  
For short times $t/\tau<<1 $,
${\cal R} \sim 1 - \sqrt{ 2 t }/ (\pi \sqrt{\tau} )$,
and for long time ${\cal R} \sim  1/2 + e^{ - t/\tau} / \pi$.
 
Using Eqs.(\ref{eq21},\ref{eqHO03},\ref{eqHO05})  the mean square displacement
of the tagged particle is 
\begin{equation}
\langle \left( x_T \right)^2 \rangle = { \pi \over N } \xi_{{\rm th}} ^2 \left[
{ 1 \over 4} - { 1 \over \pi^2} \mbox{arccot}^2 \left( \sqrt{ e^{ 2 t / \tau} - 1 } \right) \right] .
\label{eqHO06} 
\end{equation}  
For short times $t<<\tau$
\begin{equation}
\langle \left( x_T \right)^2 \rangle \sim {\xi_{{\rm th}} \over N} \sqrt{ 2 D t}
\label{eqHO07} 
\end{equation}  
Such a behavior is expected, as well known
 for short times  the diffusion process of the Ornstein Uhlenbeck process
$x\sim t^{1/2}$
 is faster
than the drift $x\sim t$  and dominates the process, i.e.
take $t<<\tau$ in Eq. (\ref{eqHO01}) and get the Green function of the
force free particle Eq. 
(\ref{eqAs01}). 
 Hence for short 
times we can neglect the
external field, but use of course the Gaussian  initial condition 
Eq. (\ref{eqHO02}).
Then the problem reduces to the short time
behavior of the Gaussian packet, free of external forces
(compare short time behavior in  Eq. 
(\ref{eqGPii}) with Eq. (\ref{eqHO07}) when $\xi\to \xi_{{\rm th}}$). 
The long time limit of Eq. 
(\ref{eqHO06}) gives 
$\langle \left( x_T \right)^2 \rangle \sim \pi \xi_{{\rm th}} ^2 / 4 N$
which is the thermal equilibrium behavior predicted more generally
in Eq. 
(\ref{eq17a}).
Behavior of the scaled  mean square displacement and its asymptotic behaviors is
presented in Fig. \ref{fig12}. 

\subsection{Thermal Initial Conditions} 

 If we assume that initially the particles are in thermal 
equilibrium, our simple formulas simplify even more. 
If $f(x_0)= 2 \exp[ - V(x_0)/k_b T]/Z$ then $r = 1/Z$. 
To see this notice that  for symmetric
potentials $V(x) = V(-x)$ we have $g(0,-x_0,t) = g(0,x_0,t)$ and
\begin{equation}
r= {2 \over Z}  \int_0 ^\infty \exp\left[ - {V(x_0) \over k_b T} \right] 
g(0,x_0, t) {\rm d} x_0 
\label{eqTI01}
\end{equation} 
yields 
\begin{equation}
r= \int_{-\infty} ^\infty { \exp\left[ - { V(x_0) \over k_b T} \right] \over Z} g(0,x_0,t) {\rm d} x_0,
\label{eqTI02}
\end{equation} 
where we assumed that the potential is binding so a stationary solution
of the Fokker-Planck equation is reached; i.e. the free particle
is excluded. Therefore $r$ [Eq. (\ref{eqTI02})] is the probability of
finding a non interacting particle at the origin, with thermal initial
conditions. Since the thermal equilibrium density is 
the stationary solution of the Fokker Planck operator, 
$r$ is time independent and equal to 
$r= \exp[ - V(0)/k_b T]/Z$. We can always take 
$V(0)= 0$ and then $r = 1/ Z$.  Examples for this behavior
are the already analyzed cases of  particles in a box with uniform initial
density [Eq. 
(\ref{eqbox02aa})] (since $Z= 2 \overline{L}$ for that case) and similarly for
particles in a harmonic potential with initial thermal density
 [Eqs.     
(\ref{eqHO02},  
\ref{eqHO03})]. 
Hence using Eq.  
(\ref{eq21})
\begin{equation}
\langle (x_T)^2 \rangle = {{\cal R} \left( 1 - {\cal R} \right) Z^2 \over 2 N}. 
\label{eqTI03}
\end{equation}
For  the symmetric and binding potentials under consideration,
 we have $\lim_{t \to \infty} {\cal R} = 1/2$ and hence Eq.
(\ref{eqTI03}) in the long time limit yields the equilibrium
behavior  Eq. 
(\ref{eq17aa}).

\subsection{Power Law Type of Initial Conditions}

 Flomenbom and Taloni \cite{Ophir}
considered particles free of external forces,
hence the free particle Green function is
\begin{equation}
g(x,x_0, t ) = { 1 \over \sqrt{ 4 \pi D t } } \exp\left[ - { \left( x - x_0 \right)^2 \over 4 D t } \right],
\label{eqOP01}
\end{equation} 
with initial conditions of power law type
\begin{equation}
f(x_0) = B | x_0 |^{- \beta}  \ \ \ \ 0<x_0 <L
\label{eqOP02}
\end{equation}
for $0<\beta<1$. Here the system size $\overline{L} \to \infty$ and
the reader should not confuse $\overline{L}$ with $L$. 
The normalization of the PDF of Eq. (\ref{eqOP02}) yields
$B= (1 - \beta) L^{\beta - 1} $. In the limit $\beta \to 0$ and $L \to \infty$
in such a way that $\rho = N/L$ remains fixed, we anticipate the
classical case of single file diffusion in the  presence of
a uniform density of particles: $\langle x^2 \rangle \propto t^{1/2}$ [Eq.
(\ref{eqbox06})]. In the opposite limit of $\beta \to 1$ the particles
are initially concentrated at the origin, and we expect the
behavior $\langle x^2 \rangle \propto t$ [Eq.   
(\ref{eqAs03a})]. Thus  $0<\beta<1$ bridges between these two known behaviors, indeed
one finds 
$\langle (x_T)^2 \rangle \propto t^{(\beta + 1 )/2}$ as shown in \cite{Ophir}.
The latter is valid for times $\sqrt{4 Dt} \ll L$ as discussed below. This indicates
that specially chosen initial condition may control the qualitative
behavior of the diffusion of the tagged particle. Here we analyze
this case using our formalism finding analytical expressions for the
mean square displacement. 

 To obtain $\langle (x_T )^2 \rangle$ all
we need to do is to find $r$ and ${\cal R}$. Using Eqs. 
(\ref{eq19},\ref{eqOP01},\ref{eqOP02}) 
\begin{equation}
r = { \left( 1 - \beta \right) \over \sqrt{\pi} } { L^{\beta-1} \over \left( \sqrt{ 4 D t } \right)^\beta } 
\int_0 ^{ {L \over \sqrt{ 4 D t}}} y^{-\beta} e^{ - y^2 } {\rm d} y,
\label{eqOP03}
\end{equation}
solving the integral
\begin{equation}
r = { \left( 1 - \beta \right) \over \sqrt{\pi} } { L^{\beta-1} \over \left( \sqrt{ 4 D t } \right)^\beta } 
{1 \over 2} \left[ \Gamma\left( {1 - \beta \over 2} \right) -
\Gamma\left({1 - \beta\over 2} , {L^2 \over 4 D t } \right)\right],
\label{eqOP03a} 
\end{equation}
where $\Gamma(a,z)$ is the incomplete Gamma function \cite{Abr}. 
The following behaviors are found for short and long
times
\begin{equation} 
r = \left\{
\begin{array}{l l}
c_r { L^{\beta - 1} \over \left( \sqrt{ 4 D t } \right)^\beta} & \ \sqrt{4 D t } \ll L \\
{ 1 \over \sqrt{ 4 \pi D t} } & \ \sqrt{4 D t} \gg L .
\end{array}
\right.
\label{eqOP04}
\end{equation}
In the short time limit we took the upper bound in integration
in Eq. (\ref{eqOP03}) to $\infty$ so
\begin{equation}
c_r = { (1 - \beta ) \over \sqrt{\pi} } \int_0 ^\infty y^{ - \beta} e^{ - y^2 } {\rm d} y = { 1 - \beta \over \sqrt{\pi} } { \Gamma\left( 1 - \beta/2 \right) \over 2}.
\label{eqOP05}
\end{equation}

Inserting Eqs. (\ref{eqOP01},\ref{eqOP02}) in Eq. (\ref{eq16}) we find
the reflection coefficient
\begin{equation}
 {\cal R} = {1 \over 2} + 
{1 \over 2 } \left( 1 - \beta \right)\left( {  L \over \sqrt{ 4 D t} } \right)^{\beta -1}
\int_0 ^{L/ \sqrt{ 4 D t}} {\rm d} y y^{- \beta} \mbox{Erf} \left( y \right) .
\label{eqOP06} 
\end{equation}
In the limit of long times we get the expected behavior $\lim_{t \to \infty} {\cal R} = 1/2$
since then half of the particles are to the left of the origin and half to the right (in statistical
sense). 
Mathematica solves the integral in the last equation
in terms of tabulated functions
\begin{equation}
\begin{array}{l}
\int_0 ^x y^{- \beta} \mbox{Erf} \left(y \right) {\rm d} y =\\
 { x^{-\beta}  \over (1 -\beta) \sqrt{\pi} } \left\{ x \left[ \sqrt{\pi} \mbox{Erf} \left( x \right) + x \mbox{E}_{\beta/2} \left( x^2 \right)\right] - x^\beta \Gamma\left( 1 - { \beta \over 2} \right) \right\}, 
\end{array}
\label{eqOP07a}
\end{equation} 
where $\mbox{E}_n (z)= \int_1^\infty {\rm d} t \exp(-z t) / t^n $ 
is the exponential integral function. 

To analyze the short time behavior we use integration by parts in Eq. 
(\ref{eqOP06}) and find
\begin{equation}
\begin{array}{c}
{\cal R} = {1 \over 2} + 
{1\over 2} \left( { L \over \sqrt{ 4 D t} } \right)^{\beta - 1}\times\\
\left[
\left( { L \over \sqrt{ 4 D t} } \right)^{1 - \beta} \mbox{Erf}\left( { L \over \sqrt{ 4 D t } } \right)
- { 2 \over \sqrt{\pi} } \int_0 ^{L/\sqrt{ 4 D t} } y^{ 1 - \beta} e^{ - y^2 } {\rm d} y  \right].
\end{array}
\label{eqOP07} 
\end{equation}
Using asymptotic properties of the $\mbox{Erf}$ function \cite{Abr}, neglecting
terms of the order of $\exp( - L^2 / 4 D t)$,  
Eq. (\ref{eqOP07})
for $\sqrt{ 4 D t } \ll L$ gives 
\begin{equation}
{\cal R} \sim 1 - c_{{\cal R}}  \left( {\sqrt{ 4 D t } \over L} \right)^{1 - \beta} ,
\label{eqOP08}
\end{equation}
with 
\begin{equation}
c_{{\cal R}} = { 1 \over \sqrt{\pi}} \int_0 ^\infty y^{1 - \beta} e^{ - y^2 } {\rm d} y ={\Gamma \left( 1 - {\beta \over 2} \right) \over 2 \sqrt{\pi} } .
\label{eqOP09}
\end{equation}

Eqs.
(\ref{eqOP03a},\ref{eqOP06},\ref{eqOP07a})  
yield the  mean square displacement
 Eq. (\ref{eq21}). The short time behavior
$\sqrt{4 D t} \ll L$ 
\begin{equation}
\langle \left( x_T \right)^2 \rangle 
\sim {1 \over 2 N} {2 \sqrt{\pi} \over (1 - \beta)^2 \Gamma\left(1 - \beta/2\right) }
L^{1-\beta} \left( \sqrt{ 4 D t } \right)^{ 1 + \beta},
\label{eqOP10}
\end{equation}  
while for long times $\sqrt{ 4 D t} \gg L$ 
\begin{equation}
\langle \left( x_{{\cal T}}\right)^2 \rangle \sim { \pi D t \over 2 N } .
\label{eqOP11}
\end{equation} 
Thus after a long time the diffusion is normal $\langle (x_T)^2 \rangle \propto t$,
exactly like  the case where all particles started
initially on the origin Eq.  
(\ref{eqAs03a}). While for short times 
$\langle (x_T)^2 \rangle \propto t^{(1 + \beta)/2}$.

Taking  the limit $\beta\to 0$ 
in Eq. (\ref{eqOP10}),
we get the
well known result of single file motion in uniform density of particles
 \cite{Ha,Le} Eq.
(\ref{eqbox06}), with the density $\rho=N/L$. 
In the opposite limit $\beta \to 1$ we have $\langle x^2 \rangle
= 
{ \pi D t \over 2 N }$ for all times
which is the same as in Eq. (\ref{eqAs03a}).  
Note that the limit $\beta\to 1$ must be treated with care,
the short time limit and the $\beta\to 1$ limit do not commute
due to the the $1/(1-\beta)^2$ divergence  in
Eq. 
(\ref{eqOP10}). When $\beta\to 1$  all particles are centered on the origin
hence we get the same behavior as in  Eq.
(\ref{eqAs03a}). 

\section{Percus Relation: Beyond Brownian Particles}
\label{SecBeyond} 

 So far we considered the case where particles are diffusing according
to the laws of Brownian motion in between collision events.
What happens for other types of motion? For example what happens
when the underlying motion itself is anomalous \cite{Ophir,Band}.

Percus \cite{Percus1} (see also \cite{Hahn,Ophir})
investigated  a general relation between the diffusion
of the tagged particle
and motion in the absence of
interactions  (free motion)
$\langle (x_T)^2 \rangle \sim \langle |x| \rangle_{{\rm free}}/\rho$.
Such a relation was suggested for normal diffusion
where we have $\langle |x| \rangle_{{\rm free}}
\propto t^{1/2}$,  so $ \langle (x_T)^2 \rangle \propto t^{1/2}$
and for a particle moving ballistically between collision
events
hence $\langle |x| \rangle_{{\rm free}} \propto t$ and
therefore  when we turn on the interactions 
$\langle (x_T)\rangle^2 \propto t$. 
This simple relation between free particle motion and the
 mean square displacement
of the  tagged interacting particle,
is expected to work for an infinite system, with a uniform
density $\rho$ of particles, and when external forces are
zero $[F(x) = 0]$. 
Here we will derive the Percus relation from our formalism, for more general
dynamics.

 Assume that the Green function of the non interacting particle
can be 
written in the scaling form 
\begin{equation}
g(x,x_0,t) = {1 \over \sqrt{K_\alpha} t^{\gamma/2} } G\left( { x - x_0 \over \sqrt{K_\gamma} t^{\gamma/2} } \right) .
\label{eqAno1}
\end{equation} 
We assume $G(y) = G(-y)$ and from normalization $\int_{-\infty} ^ \infty G(y) {\rm d} y = 1$ .
Here the free particle motion is anomalous, so that 
$\langle |x| \rangle_{{\rm free}} 
\propto t^{\gamma/2} $ for $0<\gamma$ and not equal unity. 
For example the underlying motion might be sub-diffusive continuous time
random walk (CTRW) \cite{SM,Bouchaud,Review} or fractional Brownian motion, where in the latter
case $G(y)$ is Gaussian and in the former $G(y)$ is expressed
in terms of L\'evy distributions (see some details below). 
We assume that moments of the process
are finite. The constant $K_{\gamma} $ has units of $\mbox{m}^2/\mbox{sec}^\gamma$.

 To obtain the mean square displacement of the tagged interacting
particle we calculate
the reflection coefficient
Eq. (\ref{eq16})
\begin{equation}
{\cal R} = {1 \over \overline{L}} \int_0 ^{\overline{L}} {\rm d} x_0 \int_0 ^{\overline{L}} G\left({ x- x_0 \over \sqrt{K_\gamma} t^{\gamma/2}} \right) { {\rm d} x \over 
\sqrt{K_\gamma} t^{\gamma/2} } .   
\label{eqAno2}
\end{equation}
Here we used uniform density of particle $f(x_0) = 1/ \overline{L}$,
and we will consider the limit $\overline{L} \to \infty$ with the
density $\rho$ being kept fixed. Change of variables 
$y = (x-x_0) / \sqrt{K_\gamma} t^{\gamma/2}$
and using $\overline{L} / \sqrt{K_\gamma} t^{\gamma/2} \to \infty$
we have
\begin{equation}
{\cal R} \sim {1 \over \overline{L} } \int_0 ^{\overline{L}} {\rm d} x_0 
\int_{ {- x_0 \over \sqrt{K_\gamma} t^{\gamma/2}} } ^\infty  G(y) {\rm d} y. 
\label{eqAno3}
\end{equation}
Integrating by parts, changing variables
 $y= x_0 / \sqrt{K_\gamma} t^{\gamma/2} $, and using the 
normalization
condition,  
we have 
\begin{equation}
{\cal R} \sim 
1 - {  \langle |x| \rangle_{{\rm free}} \over  2 \overline{L}} 
\label{eqAno4}
\end{equation}
where the mean of the absolute value of the free particle
motion is by its definition 
\begin{equation}
\langle |x| \rangle_{{\rm free}} = \int_{-\infty} ^\infty |x| G\left( { x \over \sqrt{K_\gamma}  t^{\gamma/2} } \right)
{{\rm d} x \over \sqrt{K_\gamma} t^{\gamma/2} }.  
\label{eqAno5}
\end{equation} 
Here we used the assumed symmetry $G(y) = G(-y)$ which means that
the underlying random walk is not biased, as a result 
$\langle |x| \rangle = \sqrt{K_\gamma} t^{\gamma/2} 2 \int_0 ^\infty y G(y) 
{\rm d} y $. It is easy to see that $r=1/2 \overline{L}$. In the limit
where $N \to \infty$ and $\overline{L} \to \infty$ we find using
Eqs. (\ref{eq21},\ref{eqAno4})   
the Percus formula for a general class of stochastic dynamics
\begin{equation}
\langle \left( x_T\right)^2 \rangle = { \langle | x| \rangle_{\rm free} \over \rho} .  
\label{eqAno6}
\end{equation}
Clearly this equation  gives
 a useful relationship between motion of a
free particle and the same particle moving in single file when it is
surrounded by identical particles whose density is $\rho$, 

 As a simple example consider fractional Brownian motion where
$g(x,x_0,t) = ( \sqrt{ 4 \pi K_\gamma t^\gamma})^{-1} \exp \left[ - (x- x_0)^2 / 4 K_\gamma t^\gamma\right]$, where $0<\gamma<2$. Then Eq. (\ref{eqAno6}) 
gives
\begin{equation}
\langle \left( x_T \right)^2 \rangle = {2 \sqrt{K_\gamma t^\gamma} \over \rho \sqrt{\pi} }  .  
\label{eqAno7}
\end{equation}
Hence if the underlying motion is ballistic $\gamma\to 2$ the motion of
the tagged particle undergoing single file dynamics is normal
with respect to time $\langle (x_T)^2 \rangle \propto t$.
For a CTRW particle in the continuum approximation the
non-interacting single particle green function is governed
by the fractional diffusion equation \cite{Schneider,MBK,Review} 
\begin{equation}
{ \partial^\gamma \over \partial t^\gamma} g(x,x_0,t) = \overline{K}_\gamma {\partial^2 \over \partial x^2} g(x,x_0,t)
\label{eqAno8}
\end{equation} 
with $0<\gamma<1$. 
From this equation it is easy to find \cite{BarkaiPRE} 
$\langle|x|\rangle_{{\rm free}} = 
\sqrt{ \overline{K}_\gamma t^\gamma} / \Gamma(1 + \gamma/2)$.
 Hence
using Eq. (\ref{eqAno6})
\begin{equation}
\langle \left( x_T \right)^2 \rangle ={ \sqrt{\overline{K}_\gamma} t^{\gamma/2} \over \rho \Gamma\left(1 + \gamma/2\right) } . 
\label{eqAno9} 
\end{equation} 
We see that independent of the mechanism of the underlying anomalous
diffusion (i.e. fractional Brownian motion, or CTRW) we get
for the tagged particle $\langle (x_{{\cal T}})^2 \rangle \sim t^{\gamma/2}$.
Further our general results show that the Green function of the
tagged particle is Gaussian even though the Green function 
of non-interacting CTRW particles  is highly non-Gaussian.

  {\bf Warning:} One should be careful  when applying our
results to the CTRW model, namely Eq. (\ref{eqAno9}) should not
be abused. 
One mechanism of anomalous diffusion are the power law waiting times
of Scher and Montroll \cite{SM}  which yield slow dynamics, as captured by the
CTRW  \cite{Bouchaud,Review} and the fractional Eq.  
(\ref{eqAno8}) \cite{BarkaiPRE}. 
Single file diffusion 
with such sub-diffusive motion as the starting point was considered
recently theoretically and with simulations in \cite{Ophir,Band}.
It should be noted that the meaning of collision in such a model
should be taken with care. For a random walk on a lattice with 
waiting times on each lattice point, one can envision several
collision rules. For example one might allow two particles
to occupy the same site at a given time, or one may consider a mechanism
were a particle once hopping into a trap already occupied will eject
the particle previously residing in the trap.  Or a particle
is allowed to jump only into an empty site, as is usually assumed.
These type of collision
rules might yield behaviors different than ours. For example
a particle stuck with a very large sojourn time, might be ejected
by another particle, hence one can imagine a situation where
some form of interaction causes the particles to move faster. 
 In our work the collision implies that we can 
let particles go past one another as if they were non interacting 
(so on a lattice two particles may occupy the same point at the same time) 
and eventually we look for the center particle.   
Even more interesting will be to investigate interacting particles
in systems with quenched disorder, since the latter,
when strong enough,
is known to lead
to non Gaussian sub-diffusion \cite{SM,Bouchaud}.  There the simple formula
Eq. (\ref{eqAno6}) is generally not expected to hold. 
Indeed in \cite{Ben} single file motion of a tagged particle in the Sinai
model was considered, the results are much richer when compared
with Eq. (\ref{eqAno6}). 

\section{Discussion}

 The one dimensional problem of motion of a tagged particle
interacting via hard core interactions with other particles
was solved using the Jepsen line. The formalism we developed
treats both Brownian and non Brownian motion in between collision
events, is suited for
rather  general external fields acting on the particles, 
for open and closed system,  and
handles also different types of initial conditions. Following
others we have mapped the problem onto a non interacting
problem using the Jepsen line. The motion of the tagged particle
belongs to the general problem of order statistics.
The problem reduces to considering a list of $2 N +1 $ random
variables and finding the distribution of
the  variable which has $N$ variables
smaller than it and $N$ larger corresponding to center particle
(note that  the right most particle we will have an
extreme value problem). Classical theory of order statistics deals however
with the case where all the random variables have identical distributions.
In contrast in the exclusion process under consideration, particles
have non identical distribution. Thus except for two cases:
i) all the particles initially on the same position and ii) equilibrium
state, the problem deals with non identically distributed random variables
(since the initial condition are non identical).
 While we
treated the problem of symmetric potential and symmetric initial
condition for the center particle in detail, it is left for 
future work to consider non-symmetric potential fields,
non symmetric initial conditions, and the dynamics of the particle
in the tails of the packet. We believe that the methods developed here
with some modifications can treat these cases too.

{\bf Acknowledgment} EB is supported by the Israel Science Foundation, and
RS by the NSF. We thank Ludvig Lizana and
 Tobias Ambj\"ornson for sharing with us their data in 
Fig. \ref{fig10} and for providing helpful insight. We also
thank M. Lomholt and A. Taloni for useful discussions. 

\section{Appendix A}

\begin{figure}
\begin{center}
\epsfig{figure=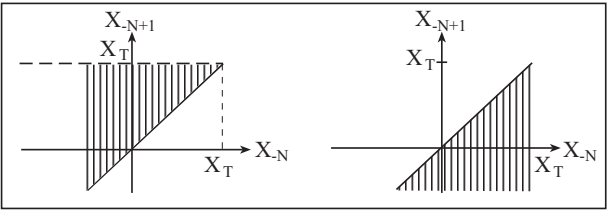, width=0.45\textwidth}
\end{center}
\caption{ 
Left panel integration in the domain
$-\infty < x_{-N} < x_{T}$, $x_{-N} < x_{-N +1}<x_T$ 
is equivalent to integration $-\infty <x_{-N} < x_T$, 
$-\infty <x_{-N +1} <x_{-N}$ (right panel).
}
\label{fig4}
\end{figure}

In this Appendix we use Boltzmann's distribution for the interacting 
system to find the PDF of the tagged center particle in equilibrium.
The multi dimensional PDF for $2 N + 1$ interacting
particles, in the presence of an external binding field $V(x)$, with $V(x) = V(-x)$, acting
on all of them is
\begin{widetext}
\begin{equation}
P\left( x_{-N}, \cdots,x_{-1} , x_{0} , x_1, \cdots, x_N \right) = 
{1 \over Z_{2 N+ 1} } \exp\left[ - \sum_{j=-N} ^N {V(x_j) \over k_b T}\right]
\theta\left(x_{-N+1} - x_{-N}\right)\theta\left( x_{-N + 2 } - x_{-N + 1} \right) \cdots
\theta\left(x_{N} - x_{N-1}\right)
\label{eqAp1}
\end{equation}
where $Z_{2 N + 1}$ is a normalizing factor, and $\theta(x)$ is a step function:
$\theta(x)=0$ if $x<0$, $\theta(x)=1$ for  $x\ge 0$.
 The  center tagged particle is $x_0=  x_T$.
To find the PDF  of $x_T$ in equilibrium, which we call
$P^{{\rm eq}}(x_T)$,  we must integrate Eq. (\ref{eqAp1})  over
all coordinates besides $x_0 \to x_T$
\begin{equation}
\begin{array}{l}
P^{{\rm eq}} \left(x_T\right) =  {\exp\left[ - { V\left( x_T \right) \over k_b T } \right] \over Z_{2 N + 1} } \times \\
 \ \\
\int_{-\infty}^{x_T} {\rm d } x_{-N} \int_{x_{-N}} ^{x_T} {\rm d} x_{-N + 1} \cdots \int_{x_{-2}} ^{x_T} {\rm d} x_{-1} 
\exp\left[ - { \sum_{j=-N} ^{-1} V\left(x_j\right) \over k_b T }  \right] 
\int_{x_{T}}^{\infty} {\rm d } x_{1} 
\int_{x_1}^{\infty} {\rm d } x_{2} 
\cdots \int_{x_{N-1} } ^\infty {\rm d} x_N 
\exp\left[ - { \sum_{j=1} ^{N} V\left(x_j\right) \over k_b T }  \right] .
\end{array}
\label{eqAp2}
\end{equation}
We rearrange the integration limits,
as explained in  
Fig. \ref{fig4}
\begin{equation}
\int_{-\infty} ^{x_T} {\rm d} x_{-N} \int_{x_{-N}} ^{x_T} {\rm d} x_{-N + 1}  \cdots = 
\int_{-\infty} ^{x_T} {\rm d} x_{-N} \int_{-\infty} ^{x_{-N} } {\rm d} x_{-N + 1}  \cdots .
\label{eqApnn}
\end{equation}
 Hence we can use
\begin{equation}
 \int_{-\infty} ^{x_T} {\rm d} x_{-N} \int_{x_{-N}} ^{x_T} {\rm d} x_{-N + 1}  \cdots = 
{1 \over 2} \left[ \int_{-\infty} ^{x_T} {\rm d} x_{-N} \int_{x_{-N}} ^{x_T} {\rm d} x_{-N + 1}  \cdots  +
\int_{-\infty} ^{x_T} {\rm d} x_{-N} \int_{-\infty} ^{x_{-N}  } {\rm d} x_{-N + 1}  \cdots \right] = {1 \over 2} \int_{-\infty} ^{x_T} {\rm d} x_N \int_{-\infty} ^{x_T} {\rm d} x_{-N + 1} \cdots .
\label{eqApnn1}
\end{equation}
Repeating this procedure  we rewrite Eq.  (\ref{eqAp2}) as
\begin{equation}
P^{{\rm eq}} \left( x_T \right) = \mbox{Nor} \left\{ \int_{-\infty} ^{x_T} {\rm d} x {\exp\left[ - {V\left( x \right) \over k_B T } \right]\over Z} \right\}^N 
 \left\{ \int_{x_T} ^{\infty} {\rm d} x {\exp\left[ - {V\left( x \right) \over k_B T } \right]\over Z}  \right\}^N 
{\exp\left[ - {V \left( x_T \right) \over k_B T} \right]\over Z}.
\label{eqAp3} 
\end{equation}
\end{widetext}
where $\mbox{Nor}$ is a normalization constant 
and $Z$ is defined in Eq. (\ref{eq13b}). 

 Thus Eq. (\ref{eqAp3}) describes a problem 
of order statistics which is extensively investigated by mathematicians, 
as mentioned  in Sec.
\ref{SecGumbel}.
 $P^{{\rm eq}}(x_T)$ 
Eq. (\ref{eqAp3}) 
is the PDF of the random variable
which has exactly $N$ random variable  larger than it and $N$ smaller.
In this sense we have transformed 
the problem to a non-interacting system, similar to the non interacting
picture in the main text. 
 The information contained in the
single non-interacting particle, i.e. the single particle
Boltzmann distribution $\exp[ - V(x)/ k_b T]/Z$ is all what is 
needed
for the calculation of the position of the tagged particle. 

  Using the  symmetry $V(x)=V(-x)$ we have
\begin{equation}
\int_{-\infty} ^{x_T} { e^{- { V(x) \over k_B T} } \over Z} {\rm d} x = { 1 \over 2} + \int_0 ^{ x_T}
 { e^{- { V(x) \over k_B T} } \over Z} {\rm d} x,
\label{eqAp4} 
\end{equation}
and
\begin{equation}
\int_{x_T} ^\infty { e^{- { V(x) \over k_B T} } \over Z} {\rm d} x = { 1 \over 2} - \int_0 ^{ x_T}
 { e^{- { V(x) \over k_B T} } \over Z} {\rm d} x.
\label{eqAp5} 
\end{equation}
Using Eqs. (\ref{eqAp4},\ref{eqAp5}), we rewrite Eq. (\ref{eqAp3}) 
\begin{equation}
\begin{array}{c}
P^{{\rm eq}} \left( x_T \right)  = \mbox{Nor} { e^{ - V(x_T)/k_b T } \over Z}e^{N \ln {1 \over 4}}  \\
 \exp\left\{ N \ln \left[ 1
- 4 \left( { \int_0 ^{x_T} e^{ - V(x)/k_b T } {\rm d} x \over Z} {\rm d} x \right)^2 \right] \right\}.
\end{array}
\label{eqAp6}
\end{equation} 
In the limit $N \to \infty$ only $x_T$ with $\int_0 ^{x_T} \exp[ - V(x)/k_b T ] {\rm d} x /Z \ll 1$
will have a measurable contribution to $P^{{ \rm eq}} (x_T)$,
since if this condition is not satisfied,
the value of $P^{{\rm eq}} (x_T)$ 
is exponentially small in $N$. Expanding the $\ln$ in Eq.
(\ref{eqAp6}) we have
\begin{equation}
P^{{\rm eq}} (x_T) \propto \exp\left[  - {V(x_T) \over k_b T } \right] \exp\left[  - 4 N \left({ \int_0 ^{x_T} {\rm d} x 
e^{ - {V(x) \over k_b T}}  \over Z} \right)^2\right].
\label{eqAp7}
\end{equation} 
Since $N>>1$ we expand $\int_0 ^{x_T} \exp( - V(x) /k_b T) {\rm d} x = x_T$ 
where we
used $V(0)=0$. 
We use  $V(x_T)/k_b T \ll N (x_T)^2 /Z^2$ which holds in the center part
of the PDF of the tagged particle (since $N>>1$, $V(0) = 0$, and
$V(x)$ is analytic) 
%
%\begin{equation}
$P^{{\rm eq}} (x_T ) \sim C \exp \left[ - { 4 N \left( x_T \right)^2 \over Z^2} \right] $
%\label{eqAp8}
%\end{equation}  
%
where $C$ is a normalization constant. 
This final approximate result is the same as Eq. 
(\ref{eq17a}), justifying the tricks used to derive our  main results.


\begin{thebibliography}{99}
\bibitem{remark0} 
For 
short times particles do not collide and then the diffusion of the tagged
particle is normal. This short time behavior is not of interest
here.  
For particles with uniform density $\rho$, and free of external
forces,  short times means
$t\ll 1/(D \rho^2)$.
\bibitem{remark}
Note that if the potential is unstable and non binding,
e.g. $V(x) = - m \omega^2 x^2 / 2$ with $m \omega^2 > 0$, the center
particle at long times can be at very large distance from its neighbors, and 
the method does not work.
\bibitem{remark3} As mentioned we treat the case where all the
particles experience the external force. In \cite{Lizana} 
the case were only the tagged particle is influenced by the force
is considered. In the latter case the relaxation to equilibrium
follows Mittag-Leffler behavior, i.e. slow relaxation
described by fractional Langevin equation. In contrast we find
that eigen values of the Fokker Planck operator describe the exponential
approach to equilibrium, and hence fractional Langevin equation
is not valid for our case. Power law relaxation describes also
the relative coordinate between two tagged particles \cite{Lizana}.
\end{thebibliography}
\end{document}